\documentclass[longauth]{aa}  

\usepackage{graphicx}

\usepackage{txfonts}
\usepackage[colorlinks=True,allcolors=blue,breaklinks=True]{hyperref}
\usepackage{natbib,twoopt}
\usepackage{placeins}

\usepackage{todonotes}
\bibpunct{(}{)}{;}{a}{}{,} 
\makeatletter
\newcommandtwoopt{\citeads}[3][][]{\href{https://ui.adsabs.harvard.edu/\#abs/#3}%
        {\def\hyper@linkstart##1##2{}%
                \let\hyper@linkend\@empty\citealp[#1][#2]{#3}}}
\newcommandtwoopt{\citepads}[3][][]{\href{https://ui.adsabs.harvard.edu/\#abs/#3}%
        {\def\hyper@linkstart##1##2{}%
                \let\hyper@linkend\@empty\citep[#1][#2]{#3}}}
\newcommandtwoopt{\citetads}[3][][]{\href{https://ui.adsabs.harvard.edu/\#abs/#3}%
        {\def\hyper@linkstart##1##2{}%
                \let\hyper@linkend\@empty\citet[#1][#2]{#3}}}
\newcommandtwoopt{\citeyearads}[3][][]%
{\href{https://ui.adsabs.harvard.edu/\#abs/#3}
        {\def\hyper@linkstart##1##2{}%
                \let\hyper@linkend\@empty\citeyear[#1][#2]{#3}}}
\makeatother

\graphicspath{{pic/}}

\newcommand{\msun}{M$_\odot$}
\newcommand{\msunyr}{\msun \ yr$^{-1}$}

\definecolor{mygreen}{RGB}{0,128,0}

\newcommand{\kms}{\mbox{km~s$^{-1}$}}

\newcommand{\mdot}{\mbox{$\dot{M}$}}
\newcommand{\my}{\mbox{$M_{\odot}$~yr$^{-1}$}}

\begin{document} 


        \title{An accreting dwarf star orbiting the S-type giant star $\pi^1$ Gru}

   \author{M.~Montargès\inst{\ref{Inst:LESIA}}
          \and
          J.~Malfait\inst{\ref{Inst:Leuven}}
          \and
          M.~Esseldeurs\inst{\ref{Inst:Leuven}}
          \and
          A.~de~Koter\inst{\ref{Inst:Amsterdam},\ref{Inst:Leuven}}
          \and
          F.~Baron\inst{\ref{Inst:GSU}}
          \and
          P.~Kervella\inst{\ref{Inst:LESIA}}
          \and
          T.~Danilovich\inst{\ref{Inst:Monash},\ref{Inst:Clayton},\ref{Inst:Leuven}}
          \and
          A.~M.~S.~Richards\inst{\ref{Inst:Manchester}}
          \and
               R.~Sahai\inst{\ref{Inst:CalTech}}
           \and           
           I.~McDonald\inst{\ref{Inst:Manchester},\ref{Inst:OpenU}}
           \and
           T.~Khouri\inst{\ref{Inst:Chalmers}}
           \and
           S.~Shetye\inst{\ref{Inst:Leuven}}
           \and
           A.~Zijlstra\inst{\ref{Inst:Manchester},\ref{Inst:Macquarie}}
               \and
               M.~Van~de~Sande\inst{\ref{Inst:Leiden}}
           \and
           I.~El~Mellah\inst{\ref{Inst:Univ_Santiago}, \ref{Inst:USACH}}
           \and
           F.~Herpin\inst{\ref{Inst:Bordeaux}}
           \and
           L.~Siess\inst{\ref{Inst:ULB}}
               \and
               S.~Etoka\inst{\ref{Inst:Manchester}}
           \and
           D.~Gobrecht\inst{\ref{Inst:Gothenburg}}
               \and
               L.~Marinho\inst{\ref{Inst:Bordeaux}}
          \and
          S.~H.~J.~Wallström~\inst{\ref{Inst:Leuven}}
         \and
         K.~T.~Wong\inst{\ref{Inst:Uppsala}}
       \and
       J.~Yates\inst{\ref{Inst:UCL}}
          }

              \institute{LIRA, Observatoire de Paris, Université PSL, Sorbonne Université, Université Paris Cité, CY Cergy Paris Université, CNRS, 92195 Meudon CEDEX, France,
              \email{\href{mailto:miguel.montarges@obspm.fr}{miguel.montarges@obspm.fr}}\label{Inst:LESIA}
              \and
              Institute of Astronomy, KU Leuven, Celestijnenlaan 200D, 3001, Leuven, Belgium\label{Inst:Leuven}
              \and
              University of Amsterdam, Anton Pannekoek Institute for Astronomy, 1090 GE, Amsterdam, The Netherlands\label{Inst:Amsterdam}
              \and
              Center for High Angular Resolution Astronomy and Department of Physics and Astronomy, Georgia State University, P.O. Box 5060, Atlanta, GA 30302-5060, USA\label{Inst:GSU}
              \and
              School of Physics \& Astronomy, Monash University, Wellington Road, Clayton 3800, Victoria, Australia\label{Inst:Monash}
              \and
              ARC Centre of Excellence for All Sky Astrophysics in 3 Dimensions (ASTRO 3D), Clayton 3800, Australia\label{Inst:Clayton}
              \and
              Jodrell Bank Centre for Astrophysics, Department of Physics and Astronomy, University of Manchester, Manchester, M13 9PL, UK\label{Inst:Manchester}
              \and
              California Institute of Technology, Jet Propulsion Laboratory, Pasadena, CA, 91109, USA\label{Inst:CalTech}
              \and
               Open University, Walton Hall, Milton Keynes, MK7 6AA, UK\label{Inst:OpenU}
              \and
              Chalmers University of Technology, Onsala Space Observatory, 43992, Onsala, Sweden\label{Inst:Chalmers}
              \and
              School of Mathematical and Physical Sciences, Macquarie University, Sydney, New South Wales, Australia\label{Inst:Macquarie}
              \and
              Leiden Observatory, Leiden University, P.O. Box 9513, 2300 RA Leiden, The Netherlands\label{Inst:Leiden}
              \and
              Departamento de Física, Universidad de Santiago de Chile, Av. Victor Jara 3659, Santiago, Chile\label{Inst:Univ_Santiago}
              \and
              Center for Interdisciplinary Research in Astrophysics and Space Exploration (CIRAS), USACH, Chile\label{Inst:USACH}
              \and
              Université de Bordeaux, Laboratoire d'Astrophysique de Bordeaux, 33615, Pessac, France\label{Inst:Bordeaux}
              \and        
              Institut d'Astronomie et d'Astrophysique, Université Libre de Bruxelles (ULB), CP 226, 1060, Brussels, Belgium\label{Inst:ULB}
              \and
              Gothenburg University\label{Inst:Gothenburg}
              \and
              Theoretical Astrophysics, Department of Physics and Astronomy, Uppsala University, Box 516, SE-751 20 Uppsala, Sweden\label{Inst:Uppsala}
            \and
            University College London, Department of Physics and Astronomy, London, WC1E 6BT, UK\label{Inst:UCL}
             }
         
   \date{Received 14 October 2024 / Accepted 23 April 2025}


 
  \abstract
   {At the end of their lives, low- to intermediate-mass stars reach the asymptotic giant branch (AGB), during which their photospheres expand by up to several hundred times and strong stellar winds develop. These changes lead to various interactions with celestial bodies in their close circumstellar environments, including mass- and angular-momentum transfer.}
   {We aim to characterize the properties of the inner companion of the S-type AGB star $\pi^1$~Gru and to identify plausible future evolutionary scenarios for this triple system.}
   {We observed $\pi^1$~Gru with the Atacama Large Millimeter/sub-millimeter Array (ALMA) and the Spectro-Polarimetric High-contrast Exoplanet REsearch (SPHERE) instrument of the Very Large Telescope (VLT), collected archival photometric data, and used the \textsc{Hipparcos}-\textit{Gaia} proper motion anomaly. We derived the best orbital parameters using Bayesian inference.}
   {In June-July 2019, the inner companion, $\pi^1$~Gru~C, was located at $37.4 \pm 2.0$~mas from the primary (a projected separation of $6.05 \pm 0.55$~au at $161.7 \pm 11.7$~pc). The best orbital solution yields a companion mass of $0.86^{+0.22}_{-0.20}$~\msun\ (using the derived mass of the primary) and a semi-major axis of $7.05^{+0.54}_{-0.57}$~au, corresponding to an orbital period of $11.0^{+1.7}_{-1.5}$~yr. The preferred solution is an elliptical orbit with eccentricity $e = 0.35^{+0.18}_{-0.17}$, although a circular orbit cannot be fully excluded. The close companion could be either a K1V$^\mathrm{F9.5V}_\mathrm{K7V}$ star or a white dwarf (WD). Ultraviolet and millimeter continuum photometry are consistent with the presence of an accretion disk around the close companion. The ultraviolet emission may originate from hot spots in an overall cooler disk, or from a hot disk if the companion is a WD.
   }
  {Although the close companion and the AGB star are interacting and an accretion disk is observed around the companion, the mass-accretion rate is too low to trigger a Type Ia supernova, but it could produce nov\ae{} every $\approx 900$~yr.  Short-wavelength, spatially resolved observations are required to further constrain the nature of the C companion. Searches for close-in companions similar to this system will improve our understanding of the physics of mass and angular momentum transfer, as well as orbital evolution during late evolutionary stages.}


   \maketitle
%


\section{Introduction\label{Sect:Intro}}

As they near the end of their existence, low- to intermediate-mass stars, i.e., stars with zero-age main sequence (ZAMS) mass (M\textsubscript{ZAMS}) lower than 8~\msun{} (where \msun{} is the solar mass) enter the asymptotic giant branch (AGB) phase. During this cool, evolved stage, the star expands significantly--from a few solar radii (R$_\odot$) to more than 1~au ($\sim 200-400$~R$_\odot$)-- and develops a slow (terminal wind velocities $v_\infty \approx 3 - 30$~km s$^{-1}$; \citeads{2018A&ARv..26....1H}) but strong stellar wind ($10^{-8}$ to $10^{-4}$~\msunyr; e.g., \citeads{2010A&A...523A..18D,2021ARA&A..59..337D}). This wind originates from the photosphere, where material is levitated by a combination of convection and pulsations. The enhanced density resulting from the dynamical motion of the surface layers allows dust grains to nucleate and grow. These grains subsequently drive and accelerate the wind through radiation pressure. Ultimately, such winds contribute significantly to the chemical enrichment of the Universe (see reviews by \citeads{2018A&ARv..26....1H} and \citeads{2021ARA&A..59..337D}). 

Early optical and infrared observations of AGB stars did not spatially resolve the close circumstellar environment. Even with the Infrared Space Observatory (ISO, \citeads{1996A&A...315L..27K}), \citetads{1997A&A...320L...1T} had to rely on spectra to characterize the molecular envelopes around these stars. The first spatially resolved studies, using speckle-masking interferometry, revealed the absence of spherical symmetry; instead, the circumstellar environments of AGB stars are inhomogeneous (or ``clumpy'') \citepads{1998A&A...333L..51W,1999A&A...346..505G}. 
Long-baseline interferometric measurements, beginning with the Infrared-Optical Telescope Array (IOTA, \citeads{1999A&A...345..221P,2004A&A...418..675P}) and the University of California Berkeley Infrared Spatial Interferometer (ISI, \citeads{1994AJ....107.1469D} and \citeads{2000ApJ...543..861M,2000ApJ...543..868M}), revealed the spatial distribution of various molecules in the circumstellar environment.
These pioneering facilities have been succeeded by the Very Large Telescope Interferometer (VLTI). Initial observations with a limited number of telescopes (thus, modest $u, v$ coverage) only permitted geometrical fitting of the interferometric observables (e.g., \citeads{2004A&A...421..703W}, \citeads{2007A&A...466.1099O}, \citeads{2008A&A...479L..21W}). In recent decades, image reconstructions have become feasible, revealing a 3D structure of the circumstellar material around AGB stars that is more complex and irregular than previously anticipated (e.g., \citeads{2015A&A...581A.127O}, \citeads{2017A&A...601A...3W}). 
In addition to these individual studies, the MID-infrared Interferometric instrument (MIDI) of the VLTI was used to conduct a systematic observation of dust around 14 AGB stars, as part of the \textit{Herschel} Mass-loss of Evolved StarS (MESS)  key program \citepads{2011A&A...526A.162G}. This MIDI large program \citepads{2017A&A...600A.136P} detected asymmetries in the dusty envelopes of five of the 14 AGB stars observed. 
At the same time, the Atacama Large Millimeter/sub-millimeter Array (ALMA) revealed asymmetries and complex structures in the gaseous component of the circumstellar environment (e.g., for R~Scl by \citeads{2012Natur.490..232M}, and R Aqr by \citeads{2018A&A...616A..61R}).

Close binaries have long been suspected of shaping the circumstellar environment and to be responsible for the various shapes of planetary nebul\ae{} (PNe; \citeads{1999ApJ...523..357M}). This hypothesis was further supported by the \textsc{Atomium} large program \citepads{2022A&A...660A..94G}, where \citetads{2020Sci...369.1497D} concluded that the rich morphology of AGB wind shapes seems most readily explained by the interaction of the AGB outflow with a close companion. Indeed, tens of such close companions have been detected and characterized around post-AGB stars (e.g., \citeads{2018A&A...620A..85O} and references therein). However, only a few have been confirmed around AGB stars, including Mira (\citeads{1997ApJ...482L.175K}), L$_2$~Puppis (\citeads{2016A&A...596A..92K}), R~Aqr (\citeads{1935ApJ....81..312M} and \citeads{2023hsa..conf..190A}), and V~Hya (\citeads{2024A&A...682A.143P}). The interaction between the wind outflow and the companion may be radiative or gravitational in nature (see observations of W~Aql by  \citeads{2024NatAs...8..308D}, and simulations from \citeads{2022MNRAS.510.1204V,2022MNRAS.513.4405A}). It may also significantly impact the chemistry of the companion, potentially revealing the binary nature of the system even after the AGB star has transitioned toward a well-hidden WD \citepads{2020A&A...639A..24E}. 

Here we focus on the S-type star $\pi^1$~Gru (161~Gru; HR~8521; HD~212087; HIP~110478). It has a parallax of $6.19 \pm 0.45$~mas \citepads{2023A&A...674A...1G}, corresponding to a distance of $161.7 \pm 11.7$~pc. In the following, we refer to the primary AGB star as $\pi^1$~Gru~A. The surface of this star has been imaged by interferometry with VLTI/PIONIER observations \citepads{2018Natur.553..310P}, revealing its granulation in exquisite detail. 
The AGB primary is known to have a distant G0V companion ($\pi^1$~Gru~B) at a projected separation of 2.8 arcsec, as reported by \citetads{1953MNRAS.113..510F}, \citetads{1992AAS...180.3708A}, and \citetads{2006AJ....132...50W}. The companion has been designated Gaia~EDR3~6518817665842868352 in \citetads{2022A&A...657A...7K}. \citetads{1992A&A...253L..33S} first mapped the CO $J=1-0$ and $2-1$ emission toward $\pi^1$~Gru using the 15-m Swedish-ESI-Submillimeter Telescope (SEST) and discovered the presence of a high-velocity bipolar outflow. He argued that $\pi^1$~Gru~B is too distant from $\pi^1$~Gru~A to cause significant interaction, and therefore is unlikely to be the origin of this outflow. 
\citetads{2006ApJ...645..605C} mapped the high-velocity outflow in CO emission with the Submillimeter Array (SMA) at higher angular resolution and suggested that it may result from interactions of the AGB wind with a second, closer companion, which may be responsible for shaping the inner AGB outflow. \citetads{2008A&A...482..561S} reinforced this hypothesis by observing asymmetries in molecular and dust shells with VLTI/MIDI. For a 2~\msun{} AGB primary, they proposed a separation between 20.8 and 74.4~mas (3.2 to 11.4~au). Using a combination of \textit{Herschel}/PACS, VLTI/AMBER, and \textsc{Hipparcos} observations, \citetads{2014A&A...570A.113M} proposed an orbital solution covering an arc of the full orbit. The position of this second companion to $\pi^1$~Gru has been directly imaged by ALMA \citepads{2020A&A...644A..61H}. \citetads{2020A&A...633A..13D} suggest that $\pi^1$ Gru experienced a mass-loss eruption 100 years ago, for about 13 yr, ejecting $4.3 \times 10^{-5}$~\msun, the origin of which has yet to be explained. This would correspond to a mass-loss rate five times higher than before the eruption. They also provide constraints on the inner companion that should exhibit an orbital period of $\sim$ 330~yr, and a separation no greater than 70~au. An eccentricity as high as 0.8 may be required to explain the branching of the spiral arms observed in the $^{12}$CO $J=3-2$ emission line.

In this article, our objective is to determine the properties of this inner companion $\pi^1$~Gru~C and its immediate ambient medium, as well as to identify possible scenarios for the future of the $\pi^{1}$~Gru system. The observations with VLT/SPHERE, ALMA and the archival photometry are presented in Sect.~\ref{Sect:Obs}. We perform data analysis in Sect.~\ref{Sect:Data_Analysis}. 
The companion characteristics and the future of the system are discussed in Sect.~\ref{Sect:Discussion}. Our concluding remarks are summarized in Sect.~\ref{Sect:Conclusion}.


\section{Observations and data reduction}\label{Sect:Obs}

\subsection{VLT/SPHERE ZIMPOL observations}

SPHERE \citepads{2019A&A...631A.155B} is the extreme adaptive optics instrument at the VLT operated by the European Southern Observatory (ESO). It is capable of observations in the visible and near-infrared domains. In this study, we used data obtained with the Zurich IMaging POLarimeter (ZIMPOL) subunit operating in the visible \citepads{2018A&A...619A...9S}. The SPHERE adaptive optics system \citepads{2006OExpr..14.7515F} allows us to get close to the diffraction limit of 17~mas in the visible (V band). Several epochs of SPHERE observations of $\pi^1$ Gru have been obtained through various filters, each preceded or followed by a spatially unresolved star acting as a flux calibrator and estimator of the point spread function (PSF). These are summarized in Table~\ref{Tab:Log_SPHERE}. The July 2019 data were acquired through the Target of Opportunity (ToO) program designed to observe the \textsc{Atomium} sources during the ALMA long baseline campaign of 2019. The ZIMPOL data reduction procedure and a general overview of these data are presented in \citetads{2023A&A...671A..96M}. The additional ZIMPOL datasets analyzed in this work were reduced (using esoreflex/SPHERE pipeline 0.45.1) and processed following the same procedure to ensure a homogeneous sample. Each observation provides four observables for each filter: the total intensity ($I$), the polarized flux ($P$), the degree of linear polarization (DoLP = $P/I$), and the polarization angle.
 
\subsection{ALMA observations\label{Sect:ALMA}}

ALMA Band 6 data of $\pi^1$ Gru were obtained during the \textsc{Atomium} large program observations. Details of the project, including the observations, data reduction, and calibration are presented in \citetads[][Table E.1]{2022A&A...660A..94G} and \citetads{2024A&A...681A..50W}. A summary of relevant parameters is provided in Table~\ref{Tab:Log_ALMA} of this article.
The ZIMPOL and ALMA extended-configuration observations are
separated by at most 15 days, as was intended with the ToO request on
VLT/SPHERE. This ensures that the data sets are contemporaneous with respect to the close circumstellar environment, given the pulsation period of $\pi^1$~Gru is $\sim 150$ days \citepads{2010A&A...523A..18D}.
This strategy provides comparable angular resolution across the spectrum at the same epoch.

\subsection{Archival optical photometry\label{SubSect:VizieR_data}}

We retrieved archival photometry from the VizieR service\footnote{\url{http://vizier.cds.unistra.fr/vizier/sed/?submitSimbad=Photometry}}, in an area of 5~arcsec around the $\pi^1$ Gru system. We manually removed obvious outliers, such as duplicated filter measurements that differed by orders of magnitude from values expected for an AGB star. Three short wavelengths (360, 450, and 550~nm) were retrieved using PySSED \citepads{2024RASTI...3...89M}. The final photometry data are compiled in Table~\ref{Tab:archive_photometry}.

We also obtained time-series photometric observations of the $\pi^1$~Gru system from the American Association of Variable Star Observers (AAVSO\footnote{\url{https://www.aavso.org}}).

\subsection{ISO spectra\label{SubSect:ISO_data}}

$\pi^1$~Gru was observed by ISO \citepads{1996A&A...315L..27K} in 1996. We retrieved the corresponding data from the ISO archive\footnote{\url{https://nida.esac.esa.int/nida-cl-web/}}.


\section{Data analysis}\label{Sect:Data_Analysis}

\subsection{Position of the companion from contemporaneous ALMA and VLT/SPHERE observations\label{SubSect:ALMA_SPHERE_pos}}

\begin{figure}
        \centering
        \includegraphics[width=\columnwidth]{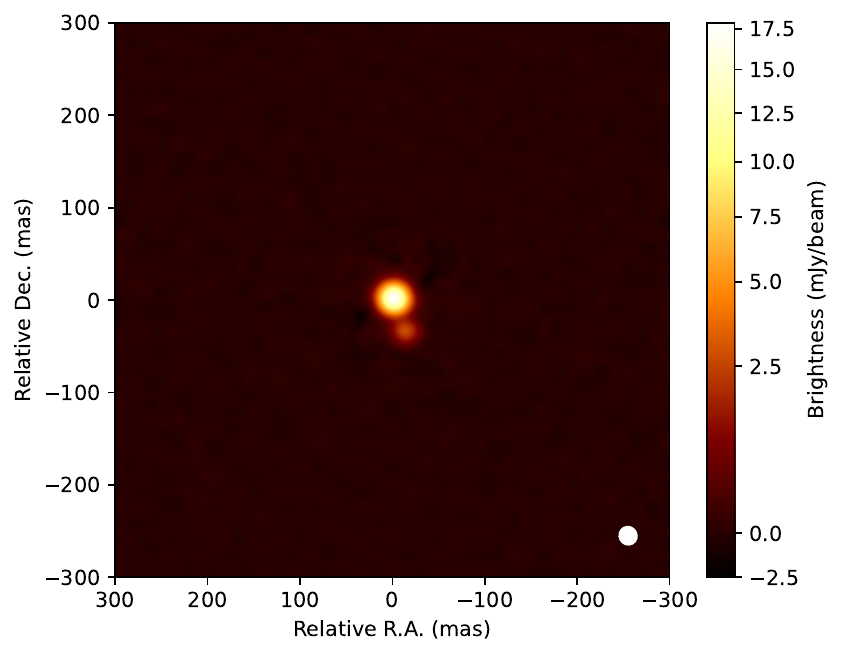}
        \caption{ALMA Band 6 continuum image of $\pi^1$~Gru~A (bright source, \citeads{2020A&A...644A..61H}) and C (faint source) from the extended configuration, observed in June-July 2019. North is up and east is to the left. The white ellipse at the bottom right corresponds to the size of the synthesized beam.}\label{Fig:ALMA_cont_high}
        
\end{figure}

Figure~\ref{Fig:ALMA_cont_high} presents the ALMA continuum map in band 6, at an angular resolution of 25~mas. The noise level, derived as the standard deviation of the noise in the map, is $\sigma = 1.31 \times 10^{-5}$~Jy beam$^{-1}.$ Two sources were spatially resolved and detected at peak values of 736$\sigma$ and 112$\sigma$, respectively. We derived their characteristics using a 2D Gaussian fit. The position of the C companion, relative to $\pi^{1}$~Gru~A, was (RA, Dec) = ($-12.3 \pm 0.1$, $-35.3 \pm 0.1$)~mas. This corresponds to an apparent separation of $37.4 \pm 0.1 \pm 2.3$~mas, or a projected separation of $6.05 \pm 0.44 \pm 0.55$~au at $161.7 \pm 11.7$~pc. The first component of the uncertainty comes from the fitting process, while the second component comes from the signal-to-noise ratio (S/N). \citetads{2020A&A...644A..61H} confirmed that the brightest component is the AGB star and the other is its suspected inner companion \citepads{2014A&A...570A.113M}. The integrated flux density over a 3$\sigma$ area of the primary was $28.3 \pm 2.8$~mJy, and $3.3 \pm 0.3$~mJy for the secondary.

\begin{figure*}
        \centering
        \includegraphics[width=\textwidth]{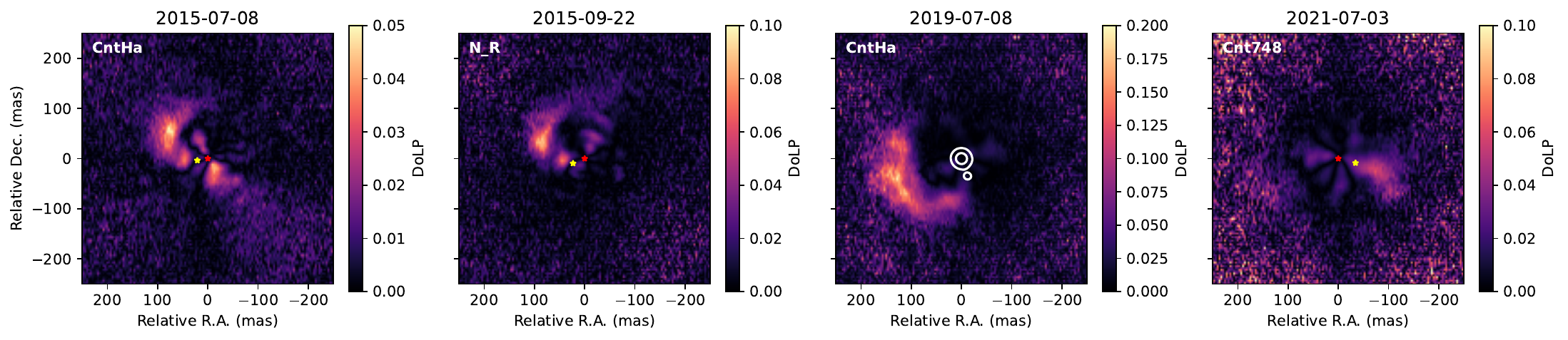}
        \caption{Degree of linear polarization in the visible measured with ZIMPOL. North is up, and east is to the left. The epoch is indicated above each image. The filter for each epoch is indicated in the top left corner of each image and was chosen to show the most prominent feature. Red stars indicate the position of the AGB star from the intensity frame and yellow stars indicate the C companion expected position at the epoch of the corresponding ZIMPOL observation, derived from the best orbit fit of Sect.~\ref{SubSect:Orbit}\label{Fig:SPHERE_DoLP_companion}. For the 2019 epoch, the 2 and 10 mJy/beam contours of the ALMA continuum image (Fig.~\ref{Fig:ALMA_cont_high}) replace these symbols. Images in all filters for each epoch are shown on Fig.~\ref{Fig:all_DoLP}.}
\end{figure*}

Figure~\ref{Fig:SPHERE_DoLP_companion} shows the degree of linear polarization (DoLP) of the ZIMPOL observations of $\pi^1$ Gru at each observation epoch, using one filter (images for all filters are shown in Fig.~\ref{Fig:all_DoLP}, and rms maps in Fig.~\ref{Fig:all_DoLP_rms}). A tail is visible in the polarized signal, most prominently from July 2015 to July 2019, and appears to rotate around the primary AGB star.  When the position of the companion from the ALMA continuum map of July 2019 is plotted onto the July 2019 ZIMPOL observation (Fig.~\ref{Fig:SPHERE_DoLP_companion}, third panel), the companion is slightly ahead of the tail, and closer to the star. We propose that this tail, which extends behind the companion and forms a spiral-like structure, results from the interaction with the stellar wind, concentrating dust and potentially accelerating its growth (similar to VLTI/MATISSE observations of V~Hya; \citeads{2024A&A...687A.306P}).
Dust is favored over gas in the polarized signal, as it is visible at multiple wavelengths \citepads{2023A&A...671A..96M}. The variation in polarized signal strength over time can be explained by changes in the angle between Earth, the tail, and the primary along the orbit. \citetads{2023A&A...671A..96M} showed that the polarized signal is best observed when dust is in the plane of the sky containing the light source. 

\subsection{Dynamics and orbital characteristics of the system \label{SubSect:Astrometry}}

In addition to the above-described position astrometry analysis, we collect various indirect observations of the binary to constrain its orbital properties. They are described in the following sections.

\subsubsection{Constraints from the gas distribution observed with ALMA\label{SubSect:ALMA_incli}}

By fitting the characteristics of the circumstellar wind from CO $J=3-2$ ALMA-ACA observations, \citetads{2017A&A...605A..28D} demonstrated that the inclination of the inner gaseous torus is $40^\circ{}$. \citetads{2020A&A...644A..61H} derived an inclination angle of $38 \pm 3^\circ{}$ using the CO ($J = 2 - 1$) higher angular resolution observations of the inner wind by the ALMA main array. In the following modeling, assuming co-planarity between the inner gaseous features and the orbital motion of the inner companion, we conservatively adopt an orbital inclination of $38 \pm 5 ^\circ{}$ with respect to the plane of the sky (with the southern part of the orbit toward Earth). We selected a broader confidence interval than that of \citetads{2020A&A...644A..61H} to enable a wider exploration of the parameter space.

\begin{figure}
        \centering
        \includegraphics[width=\columnwidth]{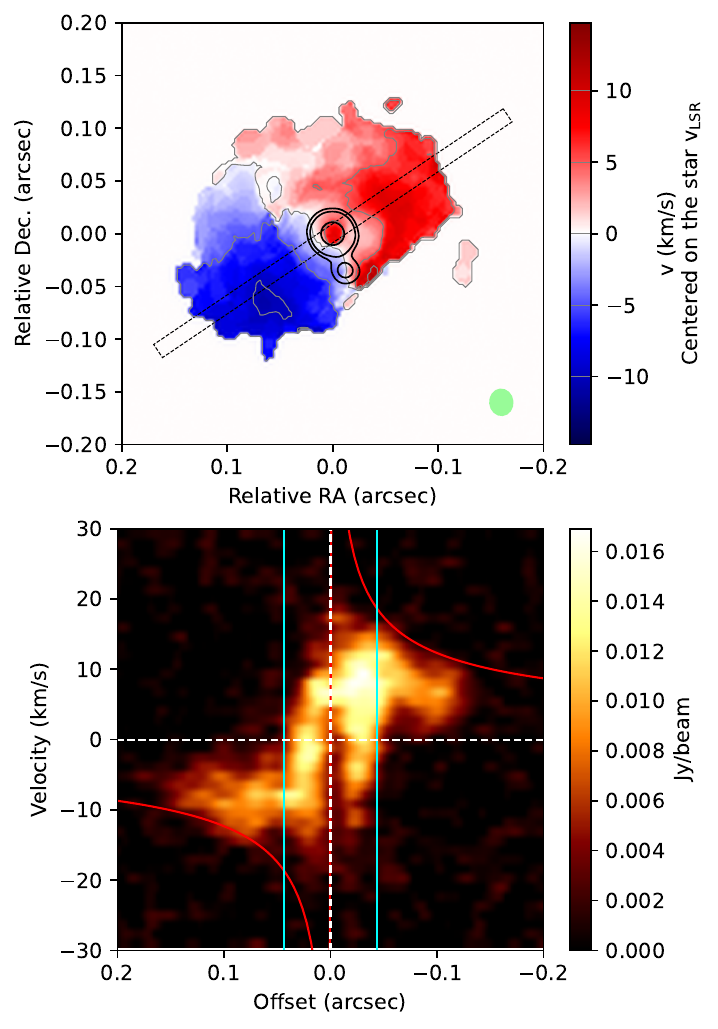}
        \caption{SiO $v=0$ $J=5-4$ emission line for $\pi^1$~Gru in our extended configuration of the ALMA array. Top panel: Moment 1 map, with black contours at 1, 2, and 10~mJy/beam indicating the continuum A and C components from Fig.~\ref{Fig:ALMA_cont_high}. The dashed rectangle represents the slit used for the P-V diagram. The green ellipse in the bottom right corner shows the ALMA beam. Bottom panel: P-V diagram with a 0.015~arcsec-wide slit oriented along a position angle (PA) of 124$^{\circ}$. The red line corresponds to the Keplerian velocity field for the central mass $M_\mathrm{A} + M_\mathrm{C}$ derived in this work, and the cyan vertical line marks the position of the C component semi-major axis.\label{Fig:PV_SiO}}
\end{figure}

\begin{figure}
        \centering
        \includegraphics[width=\columnwidth]{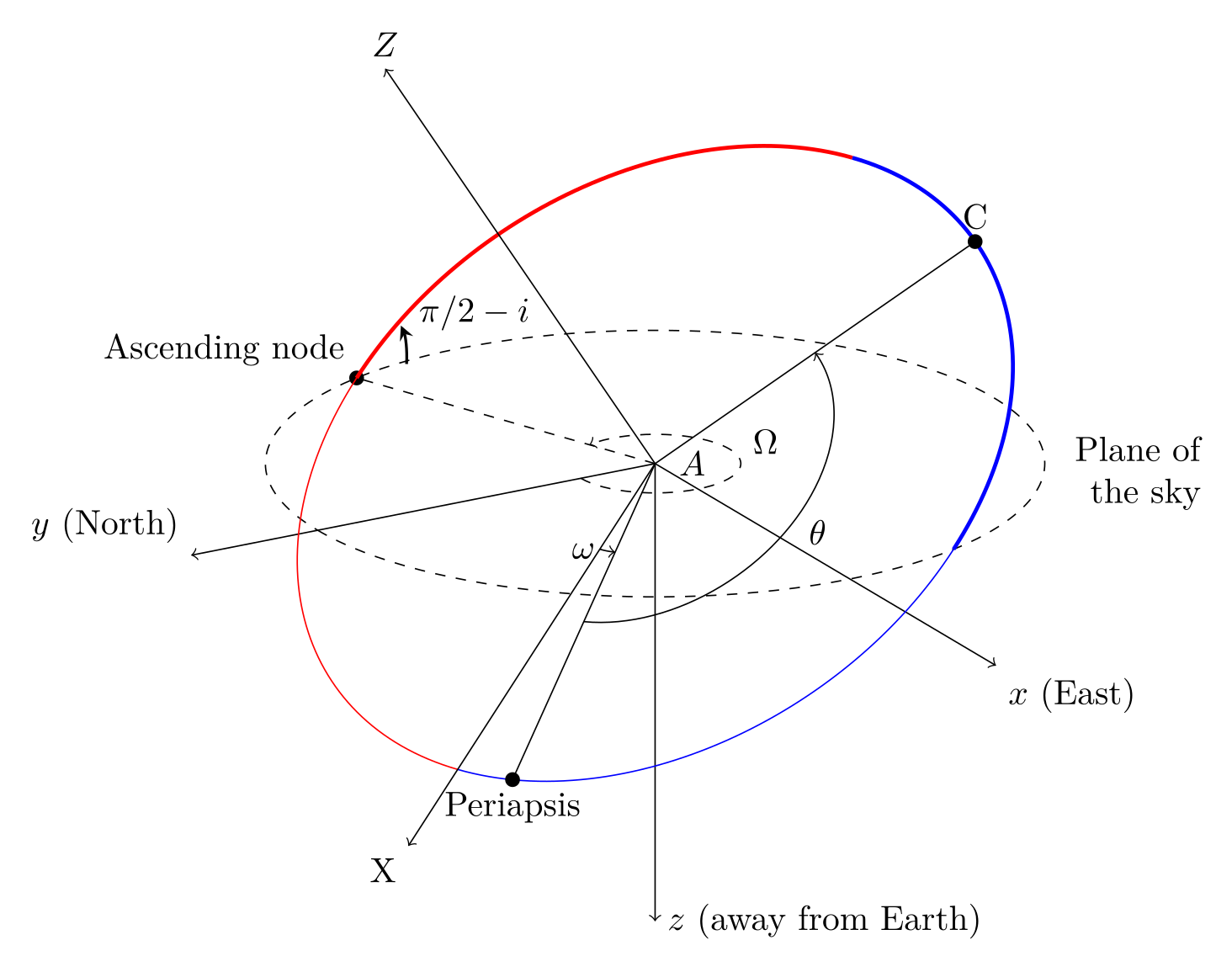}\\
        \vspace{.3cm}
        \includegraphics[width=\columnwidth]{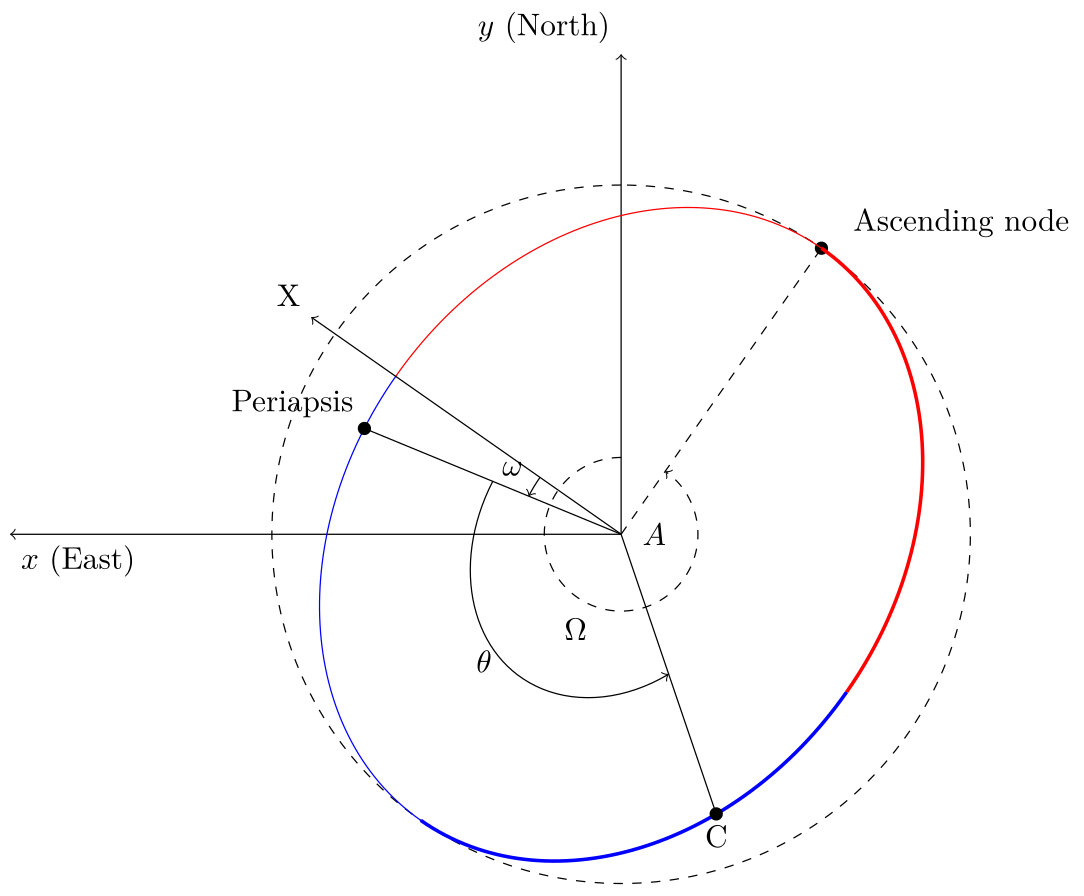}
        \caption{Sketch and parametrization of the $\pi^1$~Gru A-C system. Top panel: View down onto the orbital plane. Bottom panel: System as seen from Earth on the opposite side of the $z$ axis, with $x$ indicating the right ascension direction, and $y$ the declination. The $X$ axis lies in the plane of the orbit, and is orthogonal to the node direction. The $Z$ axis is orthogonal to the orbit plane. The orbit path above the plane of the sky (toward Earth) is shown thicker. Blue and red sections of the orbit indicate the Doppler-shift with respect to the systemic velocity of $\pi^1$~Gru~A.\label{Fig:system_sketch}}
\end{figure}

The moment 1 maps represent the mean velocity value on the line of sight at each pixel, relative to the star (assuming a local system of rest velocity for $\pi^1$ Gru of $v_\mathrm{LSR} = -12 $ km~s$^{-1}$; \citealt{2020A&A...644A..61H}), weighted by the brightness in each spectral channel. For the HCN $v = 0~J = 3 - 2$ and SiO $v = 0~J = 5 - 4$ emission lines (see Fig. 7 of \citeads{2020A&A...644A..61H} and our Fig.~\ref{Fig:PV_SiO}), the southeastern side appears blue-shifted, while the northwestern side is red-shifted. The 0 km~s\textsuperscript{-1} line-of-sight velocity direction in these moment 1 maps corresponds to the direction where the gas motion is parallel to the plane of the sky. We assume that, near the inner companion, the bulk gas motion mainly follows the orbital motion due to the changing location of the primary star caused by its orbital motion (i.e., the radial outflow is negligible). The position-velocity (P-V) diagram for the SiO $v = 0$ $J=5-4$ line in the extended ALMA configuration is shown in Fig.~\ref{Fig:PV_SiO}. It demonstrates that the maximum projected velocity of the inner gas distribution follows a Keplerian velocity field, consistent with our hypothesis. In this scenario, the direction in which the gas and companion motions lie entirely within the plane of the sky is orthogonal to the orbital node direction. The orbital node direction corresponds to the intersection of the plane of the sky and the orbital plane. Orthogonal to this orbital node direction is the $X$ axis in Fig.~\ref{Fig:system_sketch}, where the line-of-sight velocity of the companion should therefore be 0~km~s$^{-1}$ with respect to the primary. For HCN and SiO, \citetads{2020A&A...644A..61H} found the direction of minimum line-of-sight velocity of the gas of $\sim 55^\circ{}$ and $\sim 50^\circ{}$ (position angle from north to east), respectively. We adopt the mean value of $52.5 \pm 5^\circ{}$ for the null radial velocity direction of the companion, and thus a longitude of the ascending node $\Omega = 322 \pm 5^\circ{}$. The uncertainty interval corresponds to the difference between the HCN and SiO derived values.

\subsubsection{\textsc{Hipparcos} and \textit{Gaia} proper motion anomaly\label{SSect:Gaia_PMa}}

\begin{figure}
        \centering
        \includegraphics[width=\columnwidth]{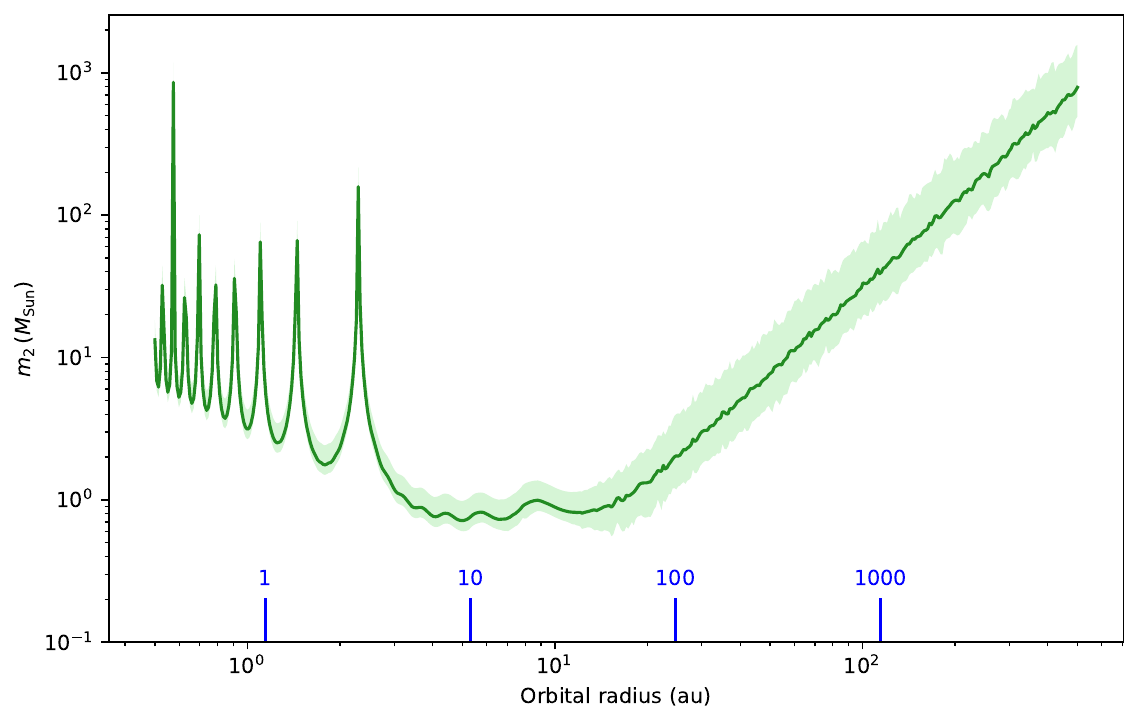}
        \caption{Proper motion anomaly sensitivity for companion detections as a function of orbital separation, for $\pi^1$~Gru, assuming a 1.5~\msun{} primary at 162~pc, and a tangential PMa of $6.07 \pm 0.33$~km~s$^{-1}$. The thick green line shows the sensitivity for a face-on system,  and the green shaded area represents the range of inclinations. Blue lines at the bottom indicate the corresponding orbital period (in years) for a circular orbit.}
        \label{Fig:PMa_sens}
\end{figure}

\citetads{2022A&A...657A...7K} performed a survey of the proper motion anomaly (PMa) for the entire \textsc{Hipparcos} catalog. The PMa is defined as the difference between the \textit{Gaia} short-term and \textsc{Hipparcos}-\textit{Gaia} long-term proper motion. We used their result to constrain the properties of the new inner companion of $\pi^1$ Gru. In this analysis, we assumed that the photocenter only tracks the primary star, as confirmed by the non-detection of the companion in the visible with VLT/SPHERE (Sect.~\ref{SubSect:SEDs}). Similar analyses have previously been performed by \citetads{2014A&A...570A.113M} using the Tycho-2 \citepads{2000A&A...355L..27H} catalog and the \textsc{Hipparcos} Intermediate Astrometric Data \citepads{2007A&A...474..653V}. The tangential velocity anomaly of the primary, measured between July 2014 and May 2017, is $6.07 \pm 0.33$~km~s$^{-1}$, with a position angle of $358.00 \pm 1.38^\circ{}$. As stated in Sect. 2.3 of \citetads{2022A&A...657A...7K}, the sensitivity function must be computed to determine what companion masses and separations can be detected. The PMa sensitivity is the result of the time smearing due to the finite observing window of both the \textsc{Hipparcos} and \textit{Gaia} missions. For a 1.5~\msun{} star at 162~pc, this sensitivity is illustrated in Fig.~\ref{Fig:PMa_sens}. As mentioned in Sect.~\ref{Sect:Intro}, $\pi^1$~Gru~A already has a known companion, a G0V star (\citeads{1953MNRAS.113..510F} and \citeads{1992AAS...180.3708A}) located at 2.8 arcsec in 2016 according to the Washington Double Star (WDS) catalog \citepads{2006AJ....132...50W}. This corresponds to a minimal physical separation of $453 \pm 33$~au. A G0V main-sequence (MS) star has a mass of 1.06~\msun{} (\citeads{2013ApJS..208....9P}, consistent with the 1.01~\msun{} estimate of \citeads{2022A&A...657A...7K}), which is far below the sensitivity limit of $\sim 400$~\msun{} at this separation (Fig.~\ref{Fig:PMa_sens}). Consequently, the contribution of the G0V companion to the PMa is negligible. The maximum sensitivity is reached in the 4-20~au domain where the inner component, designated C, is detected. We therefore attribute the PMa entirely to the new C companion.

\subsubsection{Derivation of orbital parameters\label{SubSect:Orbit}}

We used the formalism developed in \citetads[][Sect.~4.5.7 p. 79 and Sect.~4.11 p. 93]{2005ormo.book.....R} to characterize the orbit. Figure~\ref{Fig:system_sketch} illustrates the adopted parametrization of the $\pi^1$~Gru system. The $x$ and $y$ directions define the plane of the sky with $x$ corresponding to the right ascension and $y$ the declination. The $z$ axis is the line of sight away from Earth. $X$ and $Y$ define the orbital plane, with $Y$ the direction of the descending node and $X$ being orthogonal to it. The inclination $i$ is the inclination between the line of sight and the orbital plane (and corresponds to $i_\mathrm{Roy} - 90^\circ{}$, with $i_\mathrm{Roy}$ the definition of $i$ in \citeads{2005ormo.book.....R}). The longitude of the ascending node, $\Omega$, is defined as $90^\circ{} - \Omega_\mathrm{Roy}$ to match the position angle convention. The argument of the periapsis is $\omega = 90^\circ{} - \omega_\mathrm{Roy}$. The true anomaly was denoted by $\theta$. We adopted the reference frame of the primary. The direct consequence is that the tangential velocity of the C companion ($v^t_{C/A}$) is related to the \textit{Gaia} PMa of the primary $v^t_{A/\mathrm{bary}}$ through 

\begin{equation}
        v^t_{C/A} = \left( 1 + \frac{M_A}{M_C} \right) v^t_{A/\mathrm{bary}},
\end{equation} 

\noindent where $M_A$ and $M_C$ are the masses of the primary and the C companion, respectively. This simply reflects the conservation of momentum.

To solve the problem, we derived posterior probability distributions and Bayesian
evidence using the nested sampling Monte Carlo algorithm
\texttt{MLFriends} \citepads{2016S&C....26..383B,2019PASP..131j8005B} implemented in the
\texttt{UltraNest}\footnote{\url{https://johannesbuchner.github.io/UltraNest/}} package \citepads{2021JOSS....6.3001B}. Bayesian inference uses Bayes' theorem to determine the probability of hypotheses given evidence and/or observations. In other words, the prior distribution is used to derive the \textit{a posteriori} probability. We use Python's solver of the Kepler's equation developed by Dan Foreman-Mackey\footnote{\url{https://pypi.org/project/kepler.py/}}.

We explored the eight orbital parameters using the following flat prior distributions.
\begin{itemize}
        \item $M_\mathrm{A}$ between 1.1 and 1.9~\msun{} guided by a central mass estimate of 1.5~\msun{} \citepads{2014A&A...570A.113M}. The lower boundary ensures that the primary mass does not fall below the 1.1~\msun{} derived by \citet{2020A&A...644A..61H} for the C companion (and the upper boundary is chosen to be symmetric),
        \item $q= M_\mathrm{C}/M_\mathrm{A}$ between 0.2 and 0.8,
        \item $a$, the semi-major axis of the orbit, between 5 and 10~au (inferred from the separation in 2019, Sect.~\ref{SubSect:ALMA_SPHERE_pos}),
        \item $e$, the eccentricity, between 0 and 1, 
        \item $\Omega$, the longitude of the ascending node, between 0 and 360$^\circ{}$,
        \item $\omega$, the argument of the periastron, between 0 and 360$^\circ{}$, 
        \item $i$, the inclination, between 33 and 43$^\circ{}$ (Sect.~\ref{SubSect:ALMA_incli}), 
        \item $\tau_0$, the epoch of the periastron between 2009 and 2024 (deduced from a period associated with the largest considered semi-major axis).
\end{itemize}

\noindent We allowed for a full parameter range exploration of $e$, $\Omega$, and $\omega$. The fit was constrained by several observations, namely the astrometry from the ALMA continuum 2019 observation (Sect.~\ref{SubSect:ALMA_SPHERE_pos}), the 0~km~s$^{-1}$ radial velocity at PA = $52 \pm 5^\circ{}$ (Sect.~\ref{SubSect:ALMA_incli}), and the \textsc{Hipparcos}-\textit{Gaia} PMa analysis (norm and direction) in the January 2016 epoch, which corresponds to the central epoch of the \textit{Gaia} observations (Set.~\ref{SSect:Gaia_PMa}).

The uncertainty in the predictions guarantees that Bayesian fitting draws the prediction from the prior toward the observed data, resulting in a smaller in-sample error compared to the out-of-sample error, a phenomenon known as ``overfitting''. By constraining the prior to plausible values, we limit the risk of obtaining an irrelevant posterior distribution, and thus mitigate overfitting. 
We used the step sampler with 1000 live points and a number of steps equal to twice the number of parameters.

\begin{figure*}
        \centering
        \includegraphics[width=\textwidth]{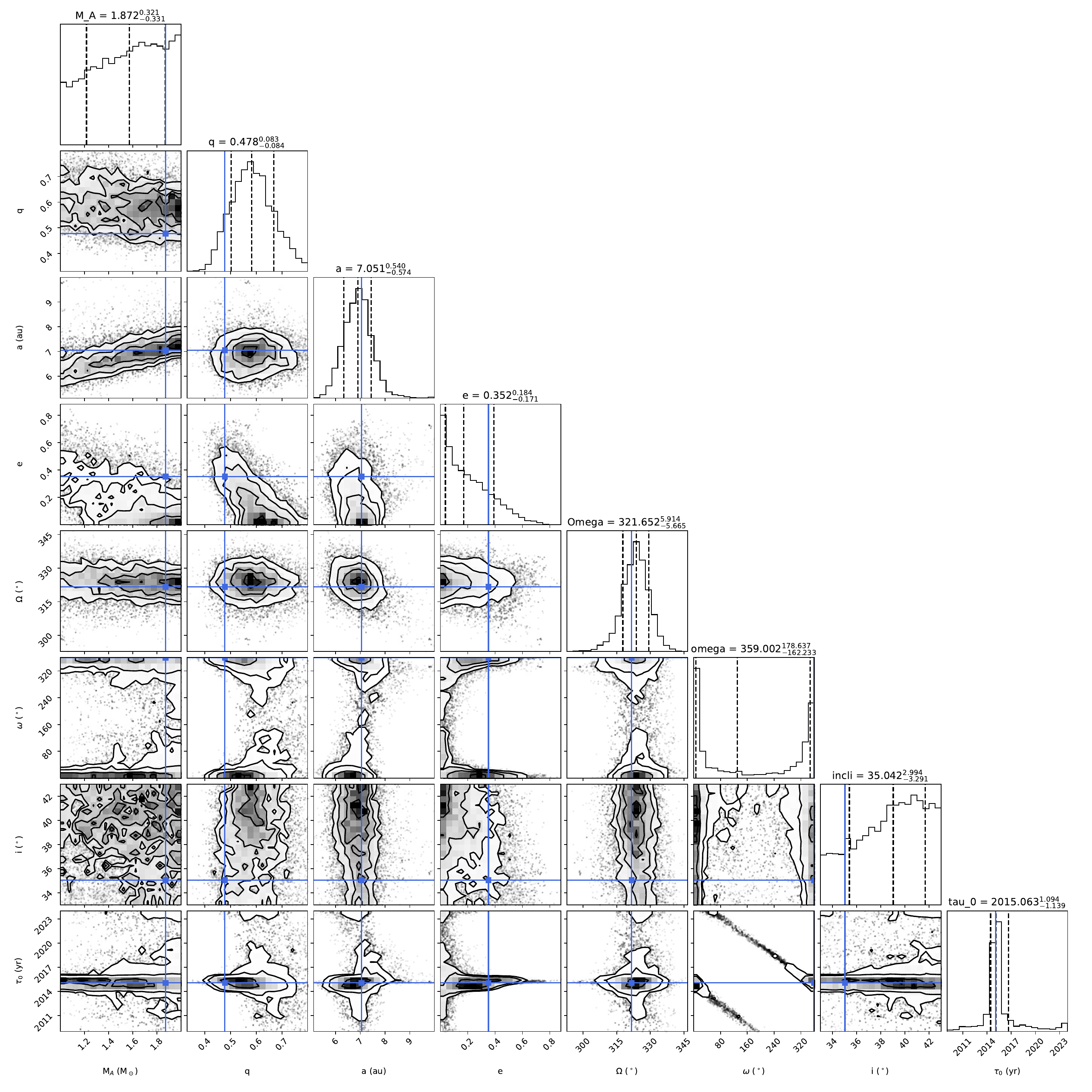}
        \caption{Corner plot of the \texttt{UltraNest} run  posterior probability distributions for the eight orbital parameters: $M_\mathrm{A}$ the mass of the AGB star in \msun, $q$ the mass ratio, $a$ the semi-major axis in au, $e$ the eccentricity, $\Omega$ the longitude of the ascending node in degrees, $\omega$ the longitude of the periastron in degrees, $i$ the inclination in degrees, and $\tau_0$ the periastron epoch. For the diagonal distributions, the central vertical dashed line indicates the mean value of the probability distribution, while the left and right vertical dashed lines correspond to the $68.3\%$ credible probability level. Blue lines indicate the maximum likelihood parameters. Contours show the 68.3\%, 39.3\% posterior and 68.3\% marginal posterior distributions. For $\Omega$ and $\omega$, the full range of angles is explored, and the horizontal axis is wrapped.\label{Fig:corner_ultranest}}
\end{figure*}

The posterior probability distribution of the parameters is shown in the corner plot in Fig.~\ref{Fig:corner_ultranest}. The best-fit parameters listed in Table~\ref{Tab:result_ultranest} correspond to the maximum likelihood, with the standard deviations obtained from the posterior probability distribution. As indicated by their relatively flat posterior distributions, the primary mass and inclination are poorly constrained. Runs with fixed values or Gaussian priors centered on literature values and their uncertainties for these parameters resulted in only marginal shifts of the other parameters within their error bars. We do not show the corner plot with these Gaussian priors as it might misleadingly suggest that the primary mass and inclination are constrained by the fitting algorithm. Using our flat priors, this would result in a C companion mass of $M_C = q . M_A = 0.86_{-0.20}^{+0.22}$~M$_\odot$ under the derived $M_A$. This value is below the upper limit of 1.1~M$_\odot$ reported by \citetads{2020A&A...644A..61H}. The orbital period corresponding to our best orbital solution is $T = 11.0^{+1.7}_{-1.5}$~yr. These values are very close to those of the smoothed particle hydrodynamics model M17 from \citetads{1999ApJ...523..357M} that predicts a spiral-shaped circumstellar environment. Such a structure has been observed around $\pi^1$~Gru by \citetads{2020A&A...633A..13D} and \citetads{2020A&A...644A..61H}. We note that the eccentricity converges toward an elliptical orbit ($e = 0.35^{+0.18}_{-0.17}$). However, the posterior distribution (Fig.~\ref{Fig:corner_ultranest}) does not allow us to rule out a circular orbit. Our estimates of the orbital period and separation are compatible with the considerations of \citetads{2008A&A...482..561S} based on VLTI/MIDI observations of the molecular and dusty circumstellar environment. From the morphology of the gaseous environment, \citetads{2020A&A...633A..13D} proposed an orbital period of 330~yr (semi-major axis of 70~au) for the C companion and an eccentricity of $\approx 0.8$. These estimates are excluded from our orbital solution. Within our best-fit solution, it is highly unlikely that the eruption they mention, which occurred 100 years ago, was caused by a direct interaction between the A and C components. Indeed, our periastron has a separation of $4.4^{+1.3}_{-1.3}$~au, i.e., $3.0^{+0.9}_{-0.9}$ times the radius of the primary \citepads{2018Natur.553..310P}. However, we cannot exclude more indirect interactions between the primary and companion near the periastron, such as the ejection of a convective plume of material by the AGB star, which triggers a localized interaction, producing an episode of enhanced mass loss.

\begin{figure}
        \centering
        \includegraphics[width=.95\columnwidth]{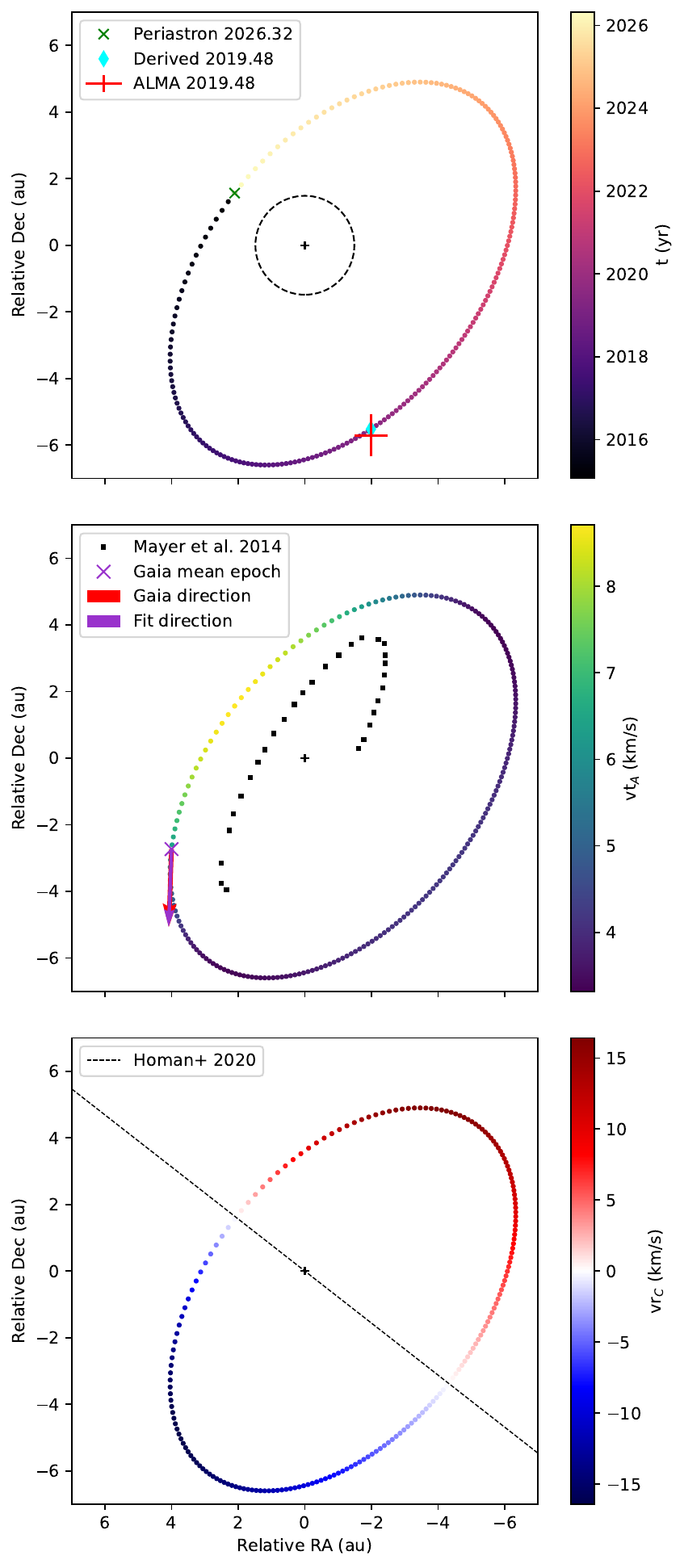}
        \caption{Best-fit orbital solution from the \texttt{UltraNest} run. Top panel: Companion position over time, showing the ALMA June 2019 detection (observation: red cross, prediction: blue diamond), and the periastron position (green cross). The black dashed circle shows the apparent size of the primary \citepads{2018Natur.553..310P}. Middle panel: Tangential velocity of the primary at the companion's orbital position, compared to the \textit{Gaia} PMa of $6.07 \pm 0.33$~km s$^{-1}$. The PMa direction (measured for A) is plotted at the position of C for January 2016, and reversed to correspond to the motion of C. Black squares indicate the orbital arc derived by \citetads{2014A&A...570A.113M}. Bottom panel: Line-of-sight velocity of the C companion. The dashed line marks the null SiO and HCN moment 1 direction identified in \citetads{2020A&A...644A..61H}.\label{Fig:orbit_ultranest}}
\end{figure}

The resulting best-fit orbit is shown in Fig.~\ref{Fig:orbit_ultranest}. We find good agreement between the 2019 ALMA detection and the model (top panel), as well as for the 0~km~s$^{-1}$ line-of-sight velocity direction (bottom panel). For the \textit{Gaia} PMa (middle panel), since we place ourselves in the reference frame of the primary, the primary itself has no motion. Nonetheless, we derived the modeled velocity of A in the barycentric frame for each epoch and plotted it at the companion's position, to verify that the value of $6.07 \pm 0.33$~km~s$^{-1}$ is reached for January 2016. We also reversed the direction of the PMa and plotted it on the orbit of the companion at the same epoch to ensure that it is tangential to the orbit. Figure~\ref{Fig:orbit_ultranest} (middle panel) shows that both constraints are met. 

In Table~\ref{Tab:TrueAnom}, we present the true anomaly for the best orbital solution at each observed epoch with both ALMA and ZIMPOL. Our best-fit model enabled us to locate the companion's position on the VLT/SPHERE-ZIMPOL images (Fig.~\ref{Fig:SPHERE_DoLP_companion}) without using them as constraints for the minimization. At each epoch where contemporaneous ALMA continuum detection is not available, the orbital solution predicts a C companion at the head of the polarized tail. This result both confirms that the dust tail is generated in the wake of the companion motion and validates the orbital solution.

\citetads{2017A&A...600A.136P} observed $\pi^1$~Gru as part of their VLTI/MIDI large program and found an elongated dusty envelope along the PA direction $140^\circ$ ($\pm 180^\circ$, \citeads{2012SPIE.8445E..1AK}). This PA is not consistent with the position of the companion relative to the primary at the corresponding epochs within our orbital solution (May 2011: PA of 277$^\circ$, September 2011: PA of 285$^\circ$), even accounting for a 180$^\circ$ degeneracy. The dusty tail observed with SPHERE is expected to trail the companion, which would imply an even smaller PA than that of the companion, and thus the inconsistency cannot be explained by a mismatch between the position of the dust tail and the companion. This discrepancy cannot be explained by the lower angular resolution of the MIDI data, which is $\sim 58$~mas (super-resolution $\lambda/2B$) and should be sufficient to resolve the dusty tail seen with SPHERE. To resolve this inconsistency, we propose that MIDI did not detect the dusty tail because it is sensitive to thermal emission from dust originating in a much larger region, which may be dominated by an asymmetric dust feature in the close circumstellar environment of $\pi^1$~Gru~A. In contrast, the polarized light images from SPHERE-ZIMPOL are preferentially sensitive to scattered light from dust that lies mainly in the plane of the sky \citepads{2023A&A...671A..96M}. However, this does not exclude the possibility that the material detected by MIDI interacts with the companion as it moves along its orbit.

\begin{table}
        \centering
        \caption{Best-fit orbital parameters of $\pi^1$~Gru~C from the \texttt{UltraNest} run.\label{Tab:result_ultranest}}
        \begin{tabular}{l l}
                \hline\hline 
                \noalign{\smallskip}
                Parameters & Values\\
                \hline
                \noalign{\smallskip}
                $M_\mathrm{A}$ & $1.54 - 2.19$~\msun\\
                \noalign{\smallskip}
                $q$ & $0.48^{+0.08}_{-0.08}$ \\
                \noalign{\smallskip}
                $a$ & $7.05^{+0.54}_{-0.57}$~au\\
                \noalign{\smallskip}
                $e$ & $0.35^{+0.18}_{-0.17}$ \\
                \noalign{\smallskip}
                $\Omega$ & $321.7^{+5.91}_{-5.67}~^\circ{}$ \\
                \noalign{\smallskip}
                $\omega$ & $359.00^{+7.87}_{-11.26}~^\circ{}$ \\
                \noalign{\smallskip}
                $i$ & $31.75 - 38.04~^\circ{}$ \\
                \noalign{\smallskip}
                $\tau_0$ & $2015.06^{+1.09}_{-1.14}$~yr \\
                \hline
        \end{tabular}
        \tablefoot{$M_\mathrm{A}$ ($M_\mathrm{C}$) : mass of the primary (close companion, resp.); q: mass ratio $M_\mathrm{C}/M_\mathrm{A}$; $a$: semi-major axis; $e$: eccentricity; $\Omega$: longitude of the ascending node; $\omega$: longitude of the periastron; $i$: inclination; $\tau_0$: epoch of periastron. Best-fit values correspond to the maximum likelihood of the posterior distribution; standard deviations correspond to the $68.3\%$ credible probability level from the posterior distribution. For $M_\mathrm{A}$ and $i$, only confidence intervals are given due to poor constraints from the \texttt{UltraNest} run. These three considerations combine with the uniform prior to produce an upper value for $M_A$ that exceeds the initial prior range.}
\end{table}

\subsection{The nature of $\pi^1$~Gru C}

We consider two scenarios for the nature of $\pi^1$~Gru~C based on the parameters derived in Sect.~\ref{SubSect:Orbit}, particularly its mass $M_\mathrm{C} = 0.86^{+0.22}_{-0.20}$~\msun. The C companion could still be an MS star, specifically a yellow dwarf. Alternatively, the companion may be a compact remnant, such as a WD. We provide a justification for both scenarios.

According to \citetads{2013ApJS..208....9P}, an MS star with the derived mass of the C component should be a yellow dwarf, corresponding to a spectral type between F9.5V and K7V. We focus on the central spectral type, K1V. Its parameters are determined by the same authors\footnote{\url{https://www.pas.rochester.edu/~emamajek/EEM_dwarf_UBVIJHK_colors_Teff.txt}} and are summarized in Table~\ref{Tab:caract_system}. 

Alternatively, the C component could be a WD. Considering the uncertainties in the mass of the C component, initial-final mass relations \citepads{2016A&A...588A..25M,2018ApJ...866...21C} suggest an initial progenitor mass as high as 3 to 4~\msun. Given the estimate of the zero-age main-sequence mass of the present-day AGB star (of about 2~\msun), it is possible that both components A and C evolved toward their present state without any interaction.
However, it is likely that this binary interaction has taken place during the system's evolution. If so, in the WD scenario, the AGB progenitor of C may have transferred mass to the then main-sequence progenitor of $\pi^{1}$~Gru~A through wind accretion and possibly Roche lobe overflow. The interaction could have reduced the mass of C more than in a single-star scenario, while increasing the mass of the current AGB star (e.g., \citeads{2017PASA...34....1D} and references therein). Such a scenario would allow for the C component to have evolved from a progenitor with a greater initial mass than in a noninteracting case (hence, with a shorter MS lifetime, allowing the WD to coexist with the present AGB).

\begin{table*}
        \centering
        \caption{Characteristics of the three components of the $\pi^1$~Gru system.}
        \label{Tab:caract_system}
        \begin{tabular}{l l l l l}
                \hline\hline
                \noalign{\smallskip}
                Parameter & A & B & C - MS scenario & C - WD scenario \\
                \hline
                \noalign{\smallskip}
                Spectral type & S5,7 (1) & G0V (2) & K1V$^\mathrm{F9.5V}_\mathrm{K7V}$ & D? \\
                \noalign{\smallskip}
                $R$ (R$_\odot$) & $319 \pm 23$ (3, 4) & 1.10 (5) & $0.80^{+0.34}_{-0.17}$ & $0.009 \pm 0.002$ \\
                $L$ (L$_\odot$) & 7440 (6) & 1.35 (5) & $0.41^{+1.10}_{-0.31}$ & Unknown \\
                \noalign{\smallskip}
                $T_{\rm eff}$ (K) & 3200 (3) & 5930 (5) & $5170^{+820}_{-1070}$ & Unknown \\
                \noalign{\smallskip}
                $M$\textsubscript{ZAMS} (M$_\odot$) & 2.0 (7) & 1.06 (5) & $0.86^{+0.22}_{-0.20}$ & $\sim 3 - 4$ \\
                \noalign{\smallskip}
                $M$\textsubscript{current} (M$_\odot$) & 1.5 (7) & 1.06 (5) & $0.86^{+0.22}_{-0.20}$ &  $0.86^{+0.22}_{-0.20}$\\
                \noalign{\smallskip}
                \hline
        \end{tabular}
        \tablefoot{$R$ and $L$ denote the radius and luminosity, respectively; $T_{\rm eff}$ is its effective temperature. For component C, the parameters are determined in this work. The MS spectral type, radius, luminosity, and effective temperature are taken from (5) using our derived mass. The WD radius is derived from (8) and its initial mass is inferred from (9).}
        \tablebib{
                (1)~\citetads{1954ApJ...120..484K};
                (2)~\citeads{1953MNRAS.113..510F} and \citeads{1992AAS...180.3708A};
                (3)~\citetads{2018Natur.553..310P};
                (4)~\citetads{2023A&A...674A...1G};
                (5)~\citetads{2013ApJS..208....9P};
                (6)~\citetads{1998A&A...329..971V};
                (7)~\citetads{2014A&A...570A.113M};
                (8)~\citetads{1972ApJ...175..417N};
                (9)~\citetads{2016A&A...588A..25M}.
        }
\end{table*} 

\subsection{Variability of the $\pi^1$ Gru system in the visible\label{SubSect:Variability}}

\begin{figure}
        \centering
        \includegraphics[width=\columnwidth]{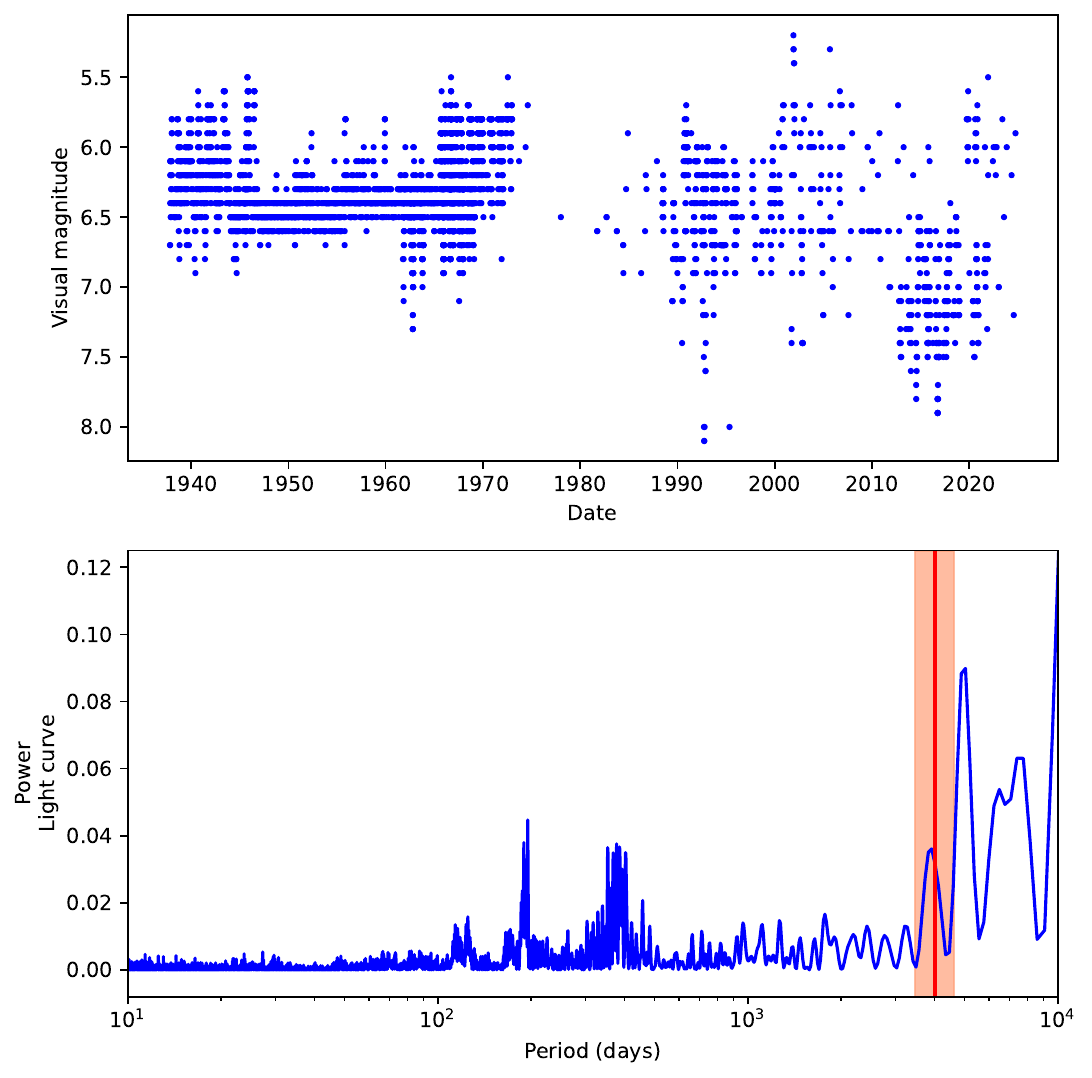}
        \caption{Variability of the $\pi^1$~Gru system. Top: Visual light curve from the AAVSO. Bottom: Periodogram (blue). The red vertical line indicates the orbital period of the best-fit \texttt{UltraNest} solution; the shaded area corresponds to the uncertainty.\label{Fig:light_curve}}
\end{figure}

The AAVSO observations of $\pi^1$~Gru began in 1937. To search for a signature of the C companion, we looked for a periodic signal at the frequency corresponding to the orbital period. We selected only the visual observations because of their better time sampling and range. We analyzed the data with the \texttt{LombScargle} \citepads{2012cidu.conf...47V,2015ApJ...812...18V} function implemented in \texttt{Astropy} \citepads{2013A&A...558A..33A,2018AJ....156..123A,2022ApJ...935..167A}. The light curve and the periodogram are shown in Fig.~\ref{Fig:light_curve}.

No significant periodic signal matching the best \texttt{UltraNest} orbital period is detected.

\subsection{Photometry from the ultraviolet to the millimeter domain: scenarios for $\pi^1$~Gru C\label{SubSect:SEDs}}

\begin{figure*}
        \centering
        \includegraphics[width=\textwidth]{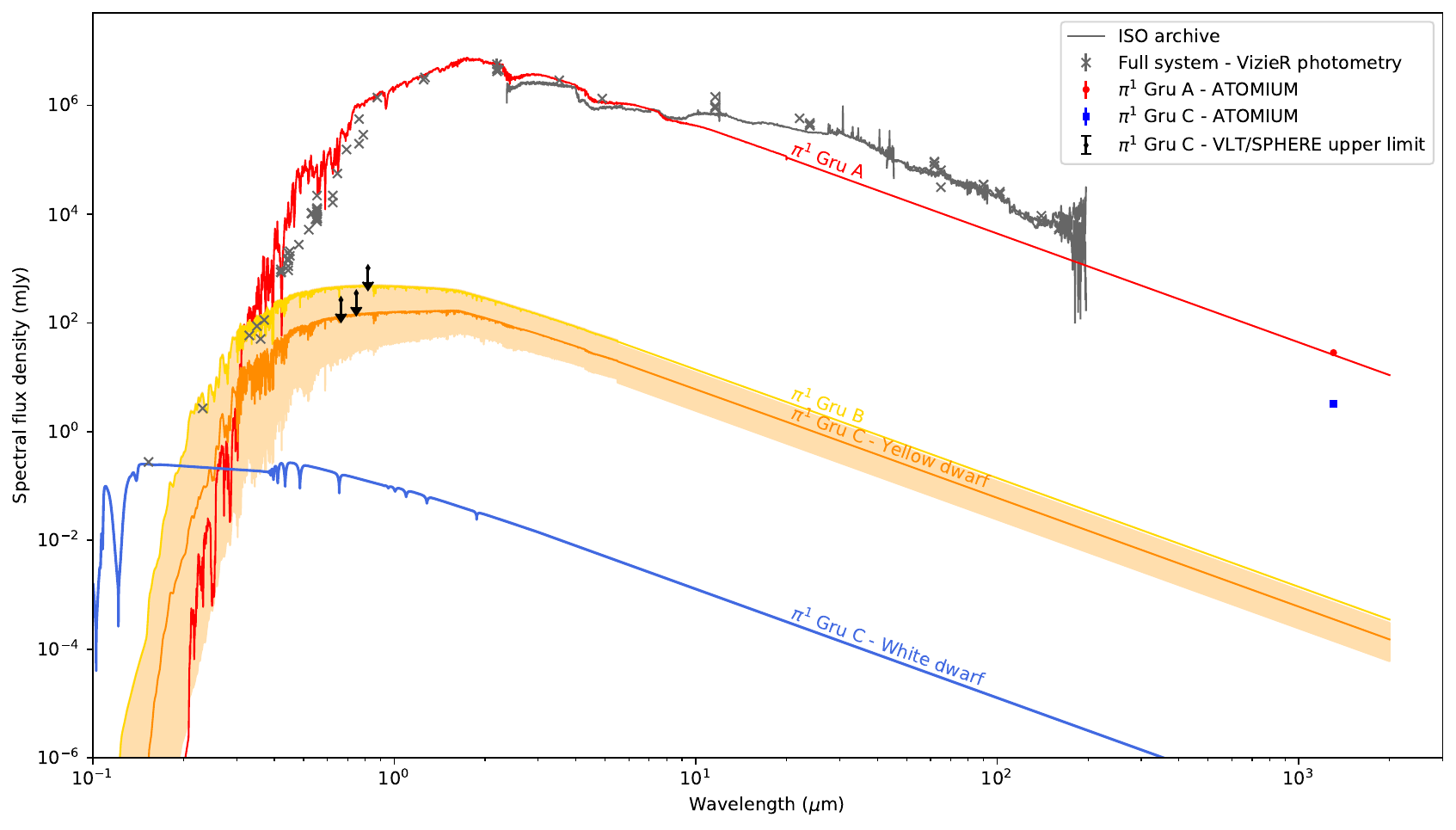}
        \caption{Photometry and spectral energy distributions (SEDs) for the $\pi^1$~Gru system. Interstellar or circumstellar reddening are neglected. The gray crosses are VizieR ultraviolet to mid-infrared photometry of the full system. The gray continuous line shows the ISO spectra. Black arrows indicate VLT/SPHERE-ZIMPOL upper detection limits of the C companion (68\% confidence). The red dot and blue square denote the integrated spectral flux density from the ALMA Band 6 continuum map for the primary and the C companion, respectively. The red curve represents a \textsc{Marcs} S-type SED for a star similar to $\pi^1$~Gru~A without circumstellar dust. The yellow and orange curves correspond to \textsc{Phoenix} SEDs for a G0V (B companion), and a K1V (C companion) MS star, respectively. The orange shading shows the uncertainty on the C companion properties under the MS star scenario. The blue curve corresponds to the WD scenario for the inner companion. For more details see Sect.~\ref{SubSect:SEDs}.}
        \label{Fig:SED}
\end{figure*}

To further investigate the nature of $\pi^1$~Gru, we used the photometry of the entire $\pi^1$~Gru system from the ultraviolet to the far-infrared, obtained from the VizieR database (Sect.~\ref{SubSect:VizieR_data} and Table \ref{Tab:archive_photometry}), which includes GALEX ultraviolet observations \citepads{2017ApJS..230...24B}. These data are plotted in Fig.~\ref{Fig:SED}. We note that ultraviolet spectra from the International Ultraviolet Explorer (IUE) satellite are also available\footnote{\url{https://dx.doi.org/10.17909/5kem-8r51}} for $\pi^1$~Gru. However, these spectra are very noisy, especially in the far-ultraviolet. Their levels are consistent with the GALEX measurements which, in contrast, have a much higher S/N. For this reason, we discard the IUE data in the following and rely only on the GALEX photometry for the ultraviolet. Fig~\ref{Fig:SED} shows also the ISO spectra described in Sect.~\ref{SubSect:ISO_data}.

\begin{figure}
        \centering
        \includegraphics[width=\columnwidth]{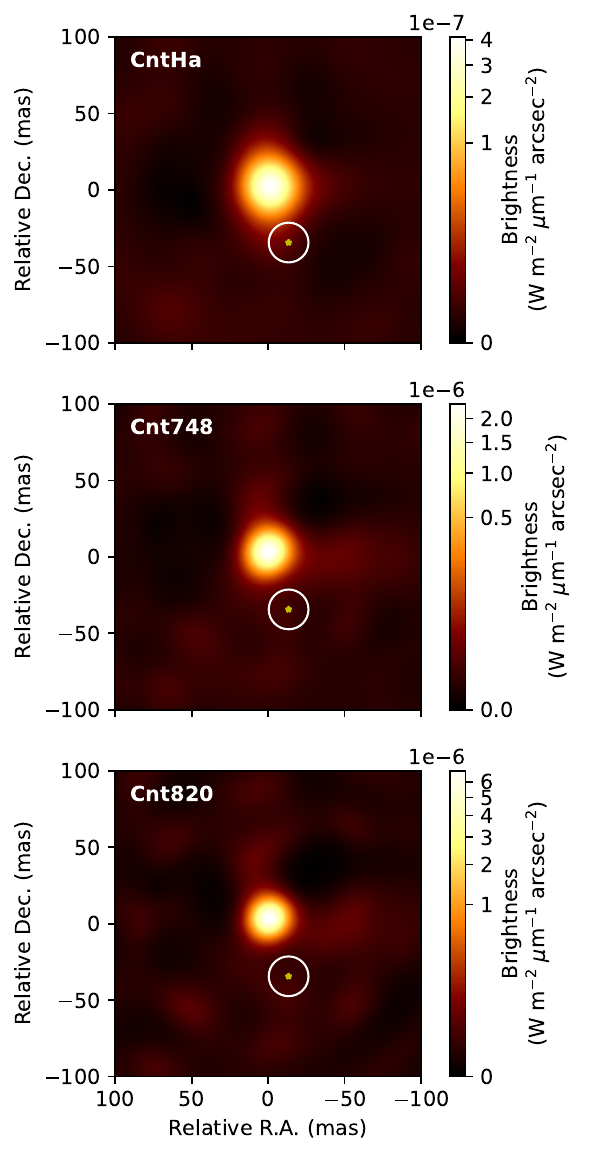}
        \caption{Deconvolved VLT/SPHERE-ZIMPOL intensity images in the visible from the July 2019 epoch. The yellow star indicates the companion position from the ALMA Band 6 continuum. The white circle indicates the PSF size (FWHM = 26.4~mas, \citeads{2023A&A...671A..96M}).}
        \label{Fig:ZIMPOL_intens_PSF}
\end{figure}

Furthermore, we deconvolved the ZIMPOL intensity observations (Stokes $I$) with the PSF images, using the \texttt{Pyraf} implementation of the Lucy-Richardson algorithm \citepads{1972JOSA...62...55R,1974AJ.....79..745L}, with 10 iterations. The inner companion remains undetected after this additional image processing (Fig.~\ref{Fig:ZIMPOL_intens_PSF}). Using the position of the companion from the contemporaneous ALMA continuum observations (Sect.~\ref{SubSect:ALMA_SPHERE_pos}), we estimated the detection limit by summing the brightness over a PSF surface area. In Fig.~\ref{Fig:SED}, we show these upper limits, along with the spectral flux density for both components from the ALMA Band 6 continuum. These limits do not account for the circumstellar material around the C companion. If it were possible to quantify this effect, the limits would likely be higher. However, providing such an estimate is not feasible due to the complex geometry of the material around $\pi^1$~Gru A and C.

To reproduce the primary component SED, we used a \textsc{Marcs} model tuned to the S-type chemistry \citepads{2017A&A...601A..10V}. To match the observed photometry, we adopted an angular diameter of 14.7~mas in place of the inferometric H band measurement of $18.37 \pm 0.18$~mas \citepads{2018Natur.553..310P}. To model $\pi^1$ Gru B and C in the yellow dwarf scenario, we retrieved \textsc{Phoenix} spectra\footnote{\url{https://phoenix.astro.physik.uni-goettingen.de/}}  \citepads{2013A&A...553A...6H} matching their characteristics from the literature and this work (Table~\ref{Tab:caract_system}). Both \textsc{Marcs} and \textsc{Phoenix} SEDs were smoothed using a rectangular window spanning 4~\AA. Beyond 5~$\mu$m, the SEDs were extrapolated using the Rayleigh-Jeans approximation. We note that both \textsc{Marcs} and \textsc{Phoenix} models do not include circumstellar dust, are static (i.e., do not consider pulsations), and assume chemical equilibrium. The SED of the WD was obtained from \citetads{2010MmSAI..81..921K}, using a radius of 0.009~R$_\odot$ inferred from the mass through the relation described by \citetads{1972ApJ...175..417N}. We selected an effective temperature of 15\,000~K matching the far-ultraviolet photometry from GALEX. 

We note ultraviolet and mid-infrared excesses from the full system photometry. Additionally, the integrated flux density of the C companion from the ALMA Band 6 continuum exceeds the expected values for either an MS yellow dwarf or a WD. These considerations are discussed in Sect.~\ref{SSect:Nature_Companion}.


\section{Discussion}\label{Sect:Discussion}

\subsection{Comparison with the 2014 orbital solution\label{SSect:Mayer2014Solution}}

\citetads{2014A&A...570A.113M} derived an orbital solution using the Tycho-2 observations combined with the \textsc{Hipparcos} Intermediate Astrometric Data. They found several solutions compatible with their dataset, with eccentricity ranging from 0.5 to 0.9 (best fit at 0.9) and orbital periods between 9.7 and 4.6 years (best fit at 6.3 years). This corresponds to a semi-major axis ranging from 6.3 to 3.5~au with best fit at 4.7~au (using their mid-interval mass ratio of 0.75). These values are just outside the range of our best orbital solution.

Using the \texttt{Dexter}\footnote{\url{http://dc.zah.uni-heidelberg.de/sdexter}} tool, we recovered their orbital plot from Fig.~7 of their paper. We note that their right ascension axis is oriented east to the right, and that they plot the photocenter displacement. We assume that the photocenter traces the motion of the AGB (see the SED in Fig.~\ref{Fig:SED}). To derive the companion's proper motion, we used the barycenter of mass definition:
\begin{equation}
        r  = - \left(1 + \frac{1}{q}\right) r_A.
\end{equation}

The resulting orbit is shown in the middle panel of Fig.~\ref{Fig:orbit_ultranest} and again shows good qualitative agreement with our solution, particularly in orientation.

\subsection{The apparently inconsistent photometry of $\pi^1$~Gru C\label{SSect:Nature_Companion}}

If $\pi^1$~Gru~C is a WD, and for most of the MS range, it is expected to escape detection from ZIMPOL visible observations. However, Fig.~\ref{Fig:SED} shows that for the warmest MS solution, the C component could be detected. Our SEDs do not take into account the potential circumstellar extinction in the $\pi^1$~Gru system, either from the AGB wind or from around the C component in particular (accretion disk). Consequently, the coolest MS solutions are preferred, although warmer solutions cannot be ruled out.

The GALEX FUV (153~nm) and NUV (231~nm) photometry retrieved from the VizieR database show that the $\pi^1$~Gru system clearly exhibits an ultraviolet excess. This excess is not necessarily a signature of binarity, as it may originate within the chromospheric emission of the AGB star \citepads{2017ApJ...841...33M}. 

We therefore investigated whether chromospheric emission from the three components in the $\pi^1$~Gru system can produce the observed NUV and FUV fluxes observed with GALEX. For components B (a main-sequence star) and C, using empirical UV-to-total luminosity ratios from \citetads{2020AJ....159..194V}, we find that while their emission (dominated by B) can explain the observed NUV flux, their combined estimated FUV flux is significantly lower. To estimate the UV chromospheric contribution from A, we apply the results from \citetads{2022Galax..10...62S}, who analyzed a large GALEX sample of AGB stars with simple chromospheric models. They showed that if the FUV/NUV flux ratio, $R_{FUV/NUV}$, is less than 0.06, the UV emission can be explained by chromospheric and photospheric emission from an AGB star. However, if $R_{FUV/NUV}$, is significantly greater than 0.06, it must arise from accretion activity. For the comibined $\pi^1$~Gru system (A+B+C), $R_{FUV/NUV}=0.1$. Given that a substantial contribution of the NUV emission likely arises from B+C, the FUV/NUV flux ratio for $\pi^1$~Gru~A is significantly greater than 0.1. This implies that the FUV emission seen toward $\pi^1$~Gru cannot be produced by chromospheric emission and instead results from accretion activity related to A and C.

The ALMA continuum imaging at the highest available angular resolution (19~mas) shows the presence of the C companion (Fig.~\ref{Fig:ALMA_cont_high}). We argue that the observed millimeter-wave flux density (blue square in Fig.~\ref{Fig:SED}) cannot originate directly from a stellar photosphere. For instance, to match the ALMA detection, we would need an M3III type star. However, this type of star (and other evolved late-type scenarios) would be well above the detection limit in the visible with ZIMPOL, which is not the case. We therefore conclude that the millimeter-wave emission toward $\pi^1$~Gru~C arises in warm material surrounding the latter which likely resides in an accretion disk. 

\subsubsection{Comparison with other close AGB binary systems\label{SSect:Other_sytems}}

In Table~\ref{Tab:Other_systems}, we compile the characteristics of other confirmed relatively close AGB binary systems.  These range from sub-stellar (L$_2$~Pup, \citeads{2016A&A...596A..92K}) to A-type MS companions (V~Hya, \citeads{2024A&A...682A.143P}). The separations span from $1.96 \pm 0.16$~au (for L$_2$~Pup, \citeads{2016A&A...596A..92K}) and 150~au (W~Aql, \citeads{2024NatAs...8..308D}), with eccentricities ranging between $0.024_{-0.017}^{+0.027}$ (V~Hya, \citeads{2024A&A...682A.143P}) to $0.91 - 0.98$ (W~Aql, \citeads{2024NatAs...8..308D}). Despite the small sample size, this parameter space covers a wide range of possibilities corresponding to those observed in post-AGB binaries \citepads{2018A&A...620A..85O}. In terms of eccentricity and separation, W~Aql \citepads{2024NatAs...8..308D} represents an extreme case, exceeding the values observed so far in post-AGB systems. This is possibly because large periods are more difficult to measure. Of these five systems, R~Aqr stand out with properties very similar to $\pi^1$~Gru. We therefore compared this system and the prototypical Mira ($o$~Ceti) to $\pi^1$~Gru.

\begin{table*}
        \caption{Characteristics of other Mira-type close binaries.\label{Tab:Other_systems}}
        \centering
        \begin{tabular}{l l l l l l}
        \hline
        \hline
        & $o$ Ceti & L$_2$ Pup & R Aqr & V Hya & W Aql \\
        \hline
        $M_1$ (\msun)  & $\sim 2$ (1) & $0.659 \pm 0.043$ (7) & $1.0 \pm 0.2$ (10) & $0.775^{+0.64}_{-0.37}$ (13)& $1.6 \pm 0.2$ (16, 17) \\
        $M_2$ (\msun)  &  $\sim 0.7$ (1) & $(1.14 \pm 1.52) \times 10^{-2}$ (7) & $0.7\pm0.2$ (10) & $2.63^{+0.63}_{-0.69}$ (13)& $1.09 - 1.04$ (16)\\
        Spec. typ. & M5-9IIIe+DA (2, 3) & M5IIIe+L/T (7, 8) & M7e+D (11)& C-N:6+A0 (14, 13) & S6/6e+ F8/G0 (18, 19) \\
        a (au) & 70-95 (4, 5) & $\geq 1.96 \pm 0.16$ (7) & $14.5 \pm 2$ (10) & $11.2^{+1.2}_{-1.5}$ (13)& $149 - 150$ (16) \\
        $e$ & $\sim 0.16$ (4) & Unknown & $0.45 \pm 0.01$ (10) &  $0.024^{+0.027}_{-0.017}$ (13) & $0.91 - 0.98$ (16)\\
        $\dot{M}$ (\msunyr) & $2.5 \times 10^{-7} $ (6) & $5 \times 10^{-7}$ (9) & $3 \times 10^{-6}$ (12) & $1.5 \times 10^{-5}$ (15) & $3.0 \times 10^{-6}$ (20) \\
        \hline  
        \end{tabular}
        \tablebib{(1)~\citet{Planesas2016};
                (2)~\citetads{1923PASP...35..323A};
                (3)~\citetads{2010ApJ...723.1188S};
                (4)~\citet{Prieur2002};
                (5)~\citetads{2018MNRAS.477.4200S};
                (6)~\citetads{2010A&A...523A..18D};
                (7)~\citetads{2016A&A...596A..92K};
                (8)~\citetads{1985ApJS...59..197B};             
                (9)~\citetads{2002MNRAS.337...79B};
                (10)~\citetads{2023hsa..conf..190A};
                (11)~\citetads{1935ApJ....81..312M};
                (12)~\citetads{2021A&A...651A...4B};
                (13)~\citetads{2024A&A...682A.143P};
                (14)~\citetads{2009ApJ...705.1298D};
                (15)~\citetads{1997A&A...326..318K};
                (16)~\citetads{2024NatAs...8..308D};
                (17)~\citetads{2020A&A...642A..20D};
                (18)~\citetads{1980ApJS...43..379K};
                (19)~\citetads{2015A&A...574A..23D};
                (20)~\citetads{2017A&A...605A.126R}.
        }
\end{table*}

The emission at the position of $o$~Ceti~B corresponds to $\sim 12$~mJy at $\sim 229$~GHz \citep{Vlemmings2015}, which is 1.7 times brighter than the emission observed for $\pi^1$~Gru~C after correcting for the smaller distance of the $o$~Ceti~AB system. At wavelengths longer than 3.1~mm, the emission is expected to be dominated by free-free emission, while at shorter wavelengths, a nearly thermal and probably compact component is expected to dominate \citep{Planesas2016}. For R~Aqr~B, free-free emission from ionized gas surrounding the WD is expected to dominate the emission at millimeter wavelengths. This is supported by the observed spectral index $\sim 1.0$ at 0.9~mm \citep{Bujarrabal2018} and the detection of the H30$\alpha$ recombination line \citep{Gomez-Garrido2024}. The base of the jets observed in H$\alpha$ with SPHERE \citep{Schmid2017} at visible wavelengths is also detected in the continuum emission at 1.3~mm \citep{Gomez-Garrido2024}. At the position of the companion, emission peaks near~5 mJy at 0.9~mm \citep{Bujarrabal2018}. If the spectral index between 0.9 and 1.3~mm is 1.0 (similar to its value at 0.9~mm), a flux density $\sim 3.5$~mJy is expected at 1.3~mm. This is 2.7 times brighter than the continuum emission of $\pi^1$~Gru~C when considering the larger distance to R~Aqr. This points toward $\pi^1$~Gru~C being a MS object, however, this remains a tentative conclusion.

\subsubsection{Considerations on the accretion disk surrounding $\pi^1$~Gru~C\label{SSect:Accretion_disk}}

Given the differing nature of the emission observed at 1.3~mm toward $o$~Ceti~B and R~Aqr~B, it is not possible to determine the source of the emission seen toward $\pi^1$~ Gru based solely on these comparisons.
To address this, we investigate the two main physical mechanisms for producing millimeter-wave continuum emission from the vicinity of C:  thermal dust emission and free-free emission from ionized gas.
The spectral slope of the emission in the millimeter range can distinguish between these two mechanisms.

We first consider thermal dust emission from an accretion disk. This disk is expected to have an outer ``hot'' spot or annulus where the primary material impacts the disk, generating shocks. Additionally, the accretion disk will have a hot inner boundary layer. If the companion has a significant magnetic field, an inner hot spot may form where material is channeled along the magnetic field. However, the rest of the disk (i.e., most of its surface area) should be significantly cooler \citepads{2016ARA&A..54..135H}. The contributions of the hot spots to the optical and mm-wave photometry are likely to be negligible.

We can compute the radial temperature distribution if the disk is optically thick and in thermal equilibrium, using the equation below \citep{Pringle1981}: 
\begin{equation}
        \sigma T_s(r)^4 = (G M / r^3) (3\,\mdot_{a} /8 \pi) [1 - (r_i/r)^{0.5}],
\end{equation}

\noindent where $\sigma$ is the Stefan-Boltzmann constant, $T_s(r)$ is the disk surface temperature at radius $r$, $r_i$ is the radius at the inner edge of the disk, $M$ is the mass of the accretor, and $\mdot_{a}$ is the rate of accretion. 
Assuming the disk emits as a blackbody, we compute the emission for various values of $\mdot_{a}$.
We computed a large grid of models with $0.001 <r_i ({\rm au}) < 0.01$. The outer disk radius was conservatively set to $r_o ({\rm au}) \le 2$ because the continuum source is unresolved with a beam of $0.025''$.  For the mass accretion rate, a typical maximum value would be $\sim$10\% of the mass-loss rate of the wind from the primary A. However, assuming an accretion rate $\mdot_{a} \le 7.7\times10^{-8}$~\my, given the mass-loss rate of A of $\mdot(A)=7.7\times10^{-7}$~\my \citepads{2020A&A...633A..13D}, we find that the resulting emission is significantly less than required. We find that the minimum accretion rate required to produce the observed 1.3~mm flux of 3.3~mJy is $\mdot_{a} \ge 3.2\times10^{-7}$~\my, which corresponds to $\sim40$\% of the primary's mass-loss rate and is therefore very unlikely.

However, noting that C is located at an average distance of about 6~au from the primary A, the mid-point of the near-side (toward A) of this disk is likely to be only about 3.5~au from the surface of A, and the dust there will also be heated by the nearby primary star, A. With $L = 7244$~L$_\odot$, $T_\mathrm{eff}=3100$~K, we find that the dust reaches a temperature of about 1300~K at a distance of 3.9 (7.5)~au from the primary for a dust emissivity $\kappa(\nu)\propto\lambda^{-p}$, where $p$ = 1.5 (0, respectively). We conclude that heating of the dust by the primary dominates over accretion, resulting in temperatures much higher than the average disk temperature due to accretion.

Assuming the dust in the accretion disk around the companion is at this temperature, then $M_d = 1.7\times10^{-7}$~(1 cm$^{2}$~g$^{-1}$/$\kappa$(1.3mm))~\msun, where $\kappa$(1.3mm) is the absolute dust emissivity at 1.3 mm. For a gas-to-dust ratio of 200, this corresponds to a gas mass of $3.4\times10^{-5}$~\msun{}. The average optical depth of this emitting region is much less than unity. The SED of this dust emission in the millimeter and submillimeter regime is expected to be $\lambda^{-(2+p)}$. 

Recent Smoothed Particle Hydrodynamics (SPH) simulations \citepads{2024A&A...691A..84M} of binary interactions between the wind of an AGB star and a low-mass companion reveal the formation of an accretion disk around the companion, with disk sizes
of 0.5-0.9~au, consistent with the sizes discussed here. The disk temperatures predicted
by these simulations can reach as high as 6000~K to 20\,000~K, considerably higher than the thermal
equilibrium situation assumed here. At such high temperatures, dust would rapidly evaporate, implying that the observed ALMA emission must originate from a different mechanism. We discuss free-free emission as a possible source below.

Free-free emission may also contribute to the observed 1.3~mm continuum emission toward $\pi^1$~Gru~C if the gas in its vicinity is ionized by the radiation field from a WD or from accretion hot spots. For example, based on the constant density model presented by \cite{Olnon1975}, a region of size $0.025''$ with an electron temperature between 3000~K and 10$^{4}$~K would produce a flux density of $\sim 3$~mJy for an electron density $\sim2\times10^{7}$~cm$^{-3}$, with a corresponding free-free optical depth (along the line of sight, measured from the center of the object) at 1.3 mm $\sim 0.02$. Given the AGB mass-loss rate $\mdot(A)=7.7\times10^{-7}$~\my and assuming a value of 5~\kms~for the expansion velocity\footnote{Since C is likely to be located within the dust radiative-acceleration zone of A's wind, we expect the latter's expansion velocity in the vicinity of C to be a modest fraction of the wind's terminal expansion velocity of 11~\kms.}, a density of about $1.5\times10^9$~cm$^{-3}$ is expected at a radius of 3.5~au. This would imply an ionization fraction $\sim 0.01$, which should be attainable in the vicinity of an ionizing source. The spectral index
of this optically thin free-free emission in the millimeter and submillimeter regime should be $\sim 0$. The detection of recombination lines would also favor this scenario.\\

The SEDs shown in Fig.~\ref{Fig:SED} do not take into account dust extinction. $\pi^1$~Gru lies outside the Galactic plane (at Galactic latitude $< -55^\circ{}$), so the interstellar dust extinction should be minimal along a line of sight of 162~pc (which is supported by the agreement between the AGB SED and the photometry for wavelengths from the V band to about 5\,$\mu$m on Fig.~\ref{Fig:SED}). According\footnote{From the \texttt{G-Tomo} website: \url{https://explore-platform.eu/sda/g-tomo/} (login required, self-registration).} to \citetads{2019A&A...625A.135L}, $\pi^1$~Gru should have interstellar extinction $A_V = 0.005$. However, the mid-infrared excess in the SED indicates the presence of significant circumstellar dust. Further investigation of the UV excess is hindered by the lack of spatially resolved information on the 3D geometry of the dust in the $\pi^1$~Gru system, particularly within the accretion disk of $\pi^1$~Gru~C. This prevents a clear discrimination between thermal dust emission and free-free emission scenarios.

\subsection{Future evolution of the system}

With such a small separation ($a = 7.05^{+0.54}_{-0.57}$~au, see Sect.~\ref{SubSect:Astrometry}) between $\pi^1$~Gru~A and C, the future evolution of the system warrants consideration. Since $\pi^1$~Gru~B lies at a separation of $453 \pm 22$~au, and assuming it does not have a high eccentricity, it is unlikely to interact, through gravitation or radiation, with the inner pair beyond some possible weak shaping of the wind \citep{Maes2021}.

If $\pi^1$~Gru~C is an MS star, it will continue to accrete material from the wind of $\pi^1$~Gru~A as the latter evolves through its AGB and post-AGB phases. This will slightly increase the mass of C and potentially alter its surface composition, producing chemical anomalies as observed in Ba dwarfs \citepads{2004agbs.book..461J}. $\pi^1$~Gru~C will contribute to shaping of the planetary nebula (PN) resulting from the demise of $\pi^1$~Gru~A. The importance of C in this process depends on its exact mass, orbital properties, and on the velocity and strength of A's stellar wind at the position of C (\citeads{2020Sci...369.1497D} and \citealt{Maes2021}). \citetads{2017A&A...605A..28D} measured a terminal wind velocity of $10 - 13$~km s$^{-1}$, and a mass-loss rate of $7.7 \times 10^{-7}$~\msunyr. However, predicting the future evolution of these parameters remains challenging. Ultimately, $\pi^1$~Gru~A will become a WD. After tens of billions of years, the presumed yellow dwarf $\pi^1$~Gru~C will evolve toward the AGB stage or, if its mass is toward the lower range, enter the hypothetical blue dwarf stage, ultimately becoming a WD \citepads{2005AN....326..913A}.

Alternatively, if $\pi^1$~Gru~C is already a WD, assuming an accretion rate of 10\% of the primary's mass-loss rate, it may produce classical nov\ae{} with a period $\approx 900$~years (see Fig.~2 of \citeads{2001ApJ...558..323H}). However, the current accretion rate is insufficient to produce a Type Ia supernova. \citetads{2020A&A...633A..13D} already concluded that the mass-loss eruption detected $\approx 100$ years ago was unlikely to have been caused by a nova event. As discussed in Sect.~\ref{SubSect:Orbit}, direct interactions between the primary and the C companion are unlikely. We propose that the enhanced mass-loss episode a century ago originated directly from the AGB, regardless of whether the C companion is an MS star or a WD.

Whether $\pi^1$~Gru~C is an MS yellow dwarf or a WD, as long as the distant B component remains gravitationally bound, the $\pi^1$~Gru system will eventually evolve into a triple WD system (similar to \object{J1953-1019}; \citealt{2019MNRAS.483..901P}). 


\section{Conclusion}\label{Sect:Conclusion}

ALMA observations of $\pi^1$~Gru have directly detected the presence of an inner companion (separation of $37.4\pm 2.0$~mas in June-July 2019). This source appears in band 6 (1.3~mm) imaging at an angular resolution of 25~mas. The detection cannot represent the photosphere of a companion star as it would correspond to an M3 giant star that should be visible in contemporaneous VLT/SPHERE-ZIMPOL adaptive optics imaging in the visible. Combining ALMA detection with \textsc{Hipparcos}-\textit{Gaia} PMa and constraints on the gas distribution, we derive an orbital solution leading to a mass of $0.86^{+0.22}_{-0.20}$~\msun{} for this new inner companion, hereafter $\pi^1$~Gru~C. The characteristics of the companion and the orbital solution are qualitatively in agreement with those derived by \citetads{2014A&A...570A.113M}, who used a more limited dataset. Our result also agrees with the upper limit of 1.1~\msun{} derived by \citetads{2020A&A...644A..61H}. The mass of companion C corresponds to that of a K1V$^\mathrm{F9.6V}_\mathrm{K7V}$ main sequence star or a WD remnant evolved from a star with a ZAMS mass of at most $\sim 3 - 4$~\msun{}, assuming a noninteracting evolutionary path (the initial mass could be higher otherwise). Photometry in ultraviolet and millimeter wavelengths indicates that this companion has to be surrounded by a locally hot accretion disk, consistent with recent hydrodynamical simulations \citepads{2024A&A...691A..84M}. The derived orbital solution is elliptical ($e = 0.35^{+0.18}_{-0.17}$), however the posterior distribution cannot totally exclude a circular solution. The corresponding orbital period we obtain is $T = 11.0^{+1.7}_{-1.5}$~yr. Both parameters lie outside the confidence interval of \citetads{2020A&A...633A..13D}, excluding the hypothesis that the eruption they estimate to have occurred a century ago was caused by a direct interaction between the C companion and the primary AGB. However, an interaction near periastron with a plume of ejected material from the AGB star cannot be ruled out. 

If the C companion is a main-sequence yellow dwarf, it will further evolve toward the AGB phase or the hypothetical blue dwarf stage, depending on its current mass, before becoming a WD.
If the C companion is a WD, its low mass does not allow it to accrete enough mass from the AGB wind to produce a type Ia supernova, although it could produce classical nova events every $\approx 900$~years.
Ultimately, within both the yellow dwarf and WD scenarios, all three components of the $\pi^1$~Gru system will end as WDs. 

Further observations of the $\pi^1$~Gru system are required to better characterize the C companion and its environment. The brightness of the AGB dominates most of the spectrum, except in the ultraviolet for the WD scenario. Investigating the C companion will require high angular resolution through the longest baselines of ALMA, or with large single-dish telescopes such as the Extremely Large Telescope (ELT), although the infrared will be excessively dominated by the AGB star. In the longer term, the Habitable Worlds Observatory may enable further study. Expanding the sample of AGB systems with close-in companions will improve constraints on the physics of tides, mass and angular momentum transfer, and orbital evolution of such binary systems.

\section*{Data availability}

The reduced VLT/SPHERE-ZIMPOL data of the $\pi^1$~Gru system are only available at the CDS via anonymous ftp to \url{cdsarc.cds.unistra.fr} (\url{130.79.128.5}) or via \url{https://cdsarc.cds.unistra.fr/viz-bin/cat/J/A+A/671/A96} and \url{https://cdsarc.cds.unistra.fr/viz-bin/cat/J/A+A/699/A22}.
The ALMA image cubes of the $\pi^1$~Gru system are available online at \url{https://almascience.eso.org/alma-data/lp/atomium}.

\begin{acknowledgements} 
    The authors acknowledge the constructive comments from the anonymous referee that considerably improved the quality of the article.
    We are thankful to Dr. Léa Planquart for indicating an error in the inclination convention.
    Based on observations collected at the European Southern Observatory under ESO programme 095.D-0397(D), 095.D-0309(B), 0103.D-0772(A), and 1104.C-0416(F).
    This paper makes use of the following ALMA data: ADS/JAO.ALMA\#2018.1.00659.L. ALMA is a partnership of ESO (representing its member states), NSF (USA) and NINS (Japan), together with NRC (Canada), MOST and ASIAA (Taiwan), and KASI (Republic of Korea), in cooperation with the Republic of Chile. The Joint ALMA Observatory is operated by ESO, AUI/NRAO and NAOJ.
    This work has made use of data from the European Space Agency (ESA) mission {\it Gaia} (\url{https://www.cosmos.esa.int/gaia}), processed by the {\it Gaia} Data Processing and Analysis Consortium (DPAC, \url{https://www.cosmos.esa.int/web/gaia/dpac/consortium}). Funding for the DPAC has been provided by national institutions, in particular the institutions participating in the {\it Gaia} Multilateral Agreement.
    We acknowledge with thanks the variable star observations from the AAVSO International Database contributed by observers worldwide and used in this research.
    This research has made use of the Washington Double Star Catalog maintained at the U.S. Naval Observatory.
    This project has received funding from the European Union's Horizon 2020 research and innovation program under the Marie Sk\l{}odowska-Curie Grant agreement No. 665501 with the research Foundation Flanders (FWO) ([PEGASUS]$^2$ Marie Curie fellowship 12U2717N awarded to M.M.). 
    MM acknowledges funding from the Programme Paris Region fellowship supported by the Région Ile-de-France. This project has received funding from the European Union’s Horizon 2020 research and innovation program under the Marie Sk\l{}odowska-Curie Grant agreement No. 945298.
    MVdS acknowledges support from the Oort Fellowship at Leiden Observatory.
    R.S.’s contribution to the research described in this publication was carried out at the Jet Propulsion Laboratory, California Institute of Technology, under a contract with NASA. R.S. thanks NASA for financial support via GALEX GO and ADAP awards. 
    A.B. acknowledges support from ANR PEPPER and ANR WATERSTARS. 
    The research leading to these results  has received funding from the European Research Council (ERC) under the European Union's Horizon 2020 research and innovation program (projects CepBin, grant agreement 695099, and UniverScale, grant agreement 951549).
    KTW acknowledges support from the ERC under the European Union's Horizon 2020 research and innovation programme (Grant agreement no. 883867, project EXWINGS).
    TD is supported in part by the Australian Research Council through a Discovery Early Career Researcher Award (DE230100183). This research was supported in part by the Australian Research Council Centre of Excellence for All Sky Astrophysics in 3 Dimensions (ASTRO 3D), through project number CE170100013.
    IM acknowledges that this project has received funding from the European Union’s Horizon 2020 research and innovation programme under grant agreement No 101004214, and acknowledges funding from UKRI/STFC through grants ST/T000414/1 and ST/X001229/1.
    SS would like to acknowledge the support of Research Foundation-Flanders (grant number: 1239522N).
    AAZ acknowledges funding from STFC under grant ST/T000414/1.
    Financial support from the Knut and Alice Wallenberg foundation is gratefully acknowledged through grant nr. KAW 2020.0081.
    This work is funded by the French National Research Agency (ANR) project PEPPER (ANR-20-CE31- 0002).
    LS is a senior research associate from F.R.S.- FNRS (Belgium).
    MS acknowledges support from NOVA.
    J.M. acknowledges support from the Research Foundation Flanders (FWO) grant G099720N.
    S.H.J.W. acknowledges support from the Research Foundation Flanders (FWO) through grant 1285221N.
    IEM acknowledges funding via the BASAL Centro de Excelencia en Astrofisica y Tecnologias Afines (CATA) grant PFB06/2007.
    This work has made use of the SPHERE Data Centre, jointly operated by OSUG/IPAG
    (Grenoble), PYTHEAS/LAM/CESAM (Marseille), OCA/Lagrange (Nice), Observatoire de Paris/LESIA
    (Paris), and Observatoire de Lyon.
    We used the SIMBAD and VIZIER databases at the CDS, Strasbourg (France, \url{http://cdsweb.u-strasbg.fr/}), and NASA's Astrophysics Data System Bibliographic Services.
    This research made use IPython \citep{PER-GRA:2007}, Numpy \citep{5725236}, Matplotlib \citep{Hunter:2007}, SciPy \citep{2020SciPy-NMeth}, Pandas \citep{reback2020pandas,mckinney-proc-scipy-2010}, Astropy (available at \url{http://www.astropy.org/}), a community-developed core Python package for Astronomy \citepads{2013A&A...558A..33A,2018AJ....156..123A,2022ApJ...935..167A}, corner \citep{corner} and  Uncertainties (available at \url{http://pythonhosted.org/uncertainties/}): a Python package for calculations with uncertainties.\\
\end{acknowledgements}


\bibliographystyle{aa}
\bibliography{pi1Gru_companion_montarges}

\begin{appendix}
        
\onecolumn
        
        
\section{Observation logs and photometry}

This appendix regroups the observation logs of the data presented in the article. In Table~\ref{Tab:Log_SPHERE} we summarize the VLT/SPHERE-ZIMPOL observations used in the analysis.

\begin{table*}[ht!]
        \centering
        \caption{\label{Tab:Log_SPHERE}VLT/SPHERE-ZIMPOL observations of $\pi^1$ Gru and its PSF calibrators.}
         \begin{tabular}{lllllll}
                \hline
                \hline
                \noalign{\smallskip}
                Date &  Time &    Object & ND & Filter 1 & Filter 2 &  Seeing \\
                &  (UT) &           &    &          &          & (arcsec)\\
                \hline
                \noalign{\smallskip}
                2015-07-08 & 02:28 & HD 121653 & OPEN & CntH$\alpha$ & N\_H$\alpha$ & 1.08 \\
                & 02:54 & HD 121653 & OPEN & TiO\_717 & Cnt748 & 1.30 \\
                & 03:16 & HD 121653 & OPEN & Cnt820 & Cnt820 & 1.13      \\
                & 04:32 & $\pi^1$ Gru & ND\_1.0 & Cnt820 & Cnt820 & 1.21\\
                & 05:11 & $\pi^1$ Gru & OPEN & CntH$\alpha$ & N\_H$\alpha$ & 1.24\\
                & 05:29 & $\pi^1$ Gru & ND\_1.0 & TiO\_717 & Cnt748 & 0.85\\
                & 05:57 & $\pi^1$ Gru & ND\_1.0 & TiO\_717 & Cnt748 & 0.73\\
                2015-09-22 & 04:40 & $\pi^1$ Gru & ND\_1.0 & V & N\_R & 1.02\\
                & 04:54 & $\pi^1$ Gru & ND\_1.0 & V & N\_R & 0.95 \\
                & 05:20 & HD 220790 & ND\_1.0 & V & N\_R & 1.01 \\
                & 05:34 & HD 220790 & ND\_1.0 & V & N\_R & 0.94 \\
                2019-07-08 & 08:47 &   $\pi^1$ Gru &   OPEN &    CntH$\alpha$ &    CntH$\alpha$ &    0.37 \\
                & 09:03 & HD 214987 &   OPEN &    CntH$\alpha$ &    CntH$\alpha$ &    0.43 \\
                & 09:20 & $\pi^1$ Gru & ND\_1.0 &   Cnt820 &   Cnt748 &    0.49 \\
                & 09:51 & HD 214987 & ND\_1.0 &   Cnt820 &   Cnt748 &    0.46 \\
                & 09:59 & HD 214987 & ND\_1.0 &   Cnt820 &   Cnt748 &    0.43 \\
                2019-09-29 & 06:09 &   $\pi^1$ Gru &   OPEN &    CntH$\alpha$ &    CntH$\alpha$ &    0.75 \\ 
                & 06:28 &   HD 214987 &   OPEN &    CntH$\alpha$ &    CntH$\alpha$ &    0.32 \\ 
                2019-10-01 & 03:00 &  $\pi^1$ Gru &   ND\_1.0 &    Cnt820 &    Cnt748 &    1.87 \\
                & 03:23 &  HD 214987 &   ND\_1.0 &    Cnt820 &    Cnt748 &    1.80 \\
                2021-07-02 & 09:06 & $\pi^1$ Gru & OPEN & CntH$\alpha$& CntH$\alpha$ & 0.47 \\
                & 09:23 & HD 214987 & OPEN & CntH$\alpha$& CntH$\alpha$ & 0.62 \\
                2021-07-03 & 07:55 & $\pi^1$ Gru & ND\_1.0 &  Cnt820 &   Cnt748 & 0.49\\
                & 08:19 & HD 214987 & ND\_1.0 &  Cnt820 &   Cnt748 & 0.25 \\
                \hline
        \end{tabular}
        \tablefoot{The characteristics of the ZIMPOL filters are available online: \url{https://www.eso.org/sci/facilities/paranal/instruments/sphere/inst/filters.html}}
\end{table*}

\FloatBarrier

\noindent Table~\ref{Tab:Log_ALMA} contains the simplified log of the ALMA observations. 

\begin{table}[ht!]
        \caption{ALMA \textsc{Atomium} observations of $\pi^1$~Gru. \label{Tab:Log_ALMA}}
        \centering
        \begin{tabular}{c c c }
                \hline\hline
                \noalign{\smallskip}
                Date & Array configuration & Beam size (arcsec) \\
                \hline
                \noalign{\smallskip}
                2018-10-28 & Intermediate & $0.24 - 0.25$ \\
                2018-10-31 & Intermediate & $0.24 - 0.25$ \\
                2018-12-25 & Compact &  $0.77 - 0.87$ \\
                2019-03-19 & Compact & $0.77 - 0.87$ \\
                2019-06-23 & Extended & 0.019 \\
                2019-07-06 & Extended & 0.019 \\
                \hline
        \end{tabular}
        \tablefoot{The extended configuration corresponds to ALMA configurations C43-8/C43-9, the intermediate groups C43-5/C43-6, and the compact configuration is the C43-2.}
\end{table}

\FloatBarrier

\noindent Finally, in Table~\ref{Tab:archive_photometry}, we list the archival photometry data used to characterize the $\pi^1$~Gru system, as well as their origin.

\begin{table}[ht!]
        \caption{\label{Tab:archive_photometry}Archival photometry of the $\pi^1$ Gru system.}
        \centering
        \begin{tabular}{lll}    
                \hline\hline
                Wavelength ($\mu$m) & Photometry (mJy) & References \\
                \hline
                0.55 & $1.29 \times 10^{4}$ & (1) \\
                0.53 & $1.06 \times 10^{4}$ &  \\
                0.42 & $(9.63 \pm 0.12) \times 10^{2}$ &  \\
                101.95 & $2.33 \times 10^{4}$ & (2) \\
                61.85 & $7.73 \times 10^{4}$ &  \\
                23.88 & $4.37 \times 10^{5}$ &  \\
                11.59 & $9.08 \times 10^{5}$ &  \\
                0.53 & $1.03 \times 10^{4}$ & (3) \\
                0.42 & $8.55 \times 10^{2}$ &  \\
                0.53 & $(1.03 \pm 0.01) \times 10^{4}$ & (4) \\
                0.42 & $(8.55 \pm 0.16) \times 10^{2}$ &  \\
                0.44 & $(9.23 \pm 0.17) \times 10^{2}$ & (5) \\
                0.55 & $9.84 \times 10^{3}$ & (6) \\
                0.44 & $1.48 \times 10^{3}$ &  \\
                0.53 & $(1.03 \pm 0.01) \times 10^{4}$ & (7) \\
                0.42 & $(8.56 \pm 0.14) \times 10^{2}$ &  \\
                0.79 & $2.87 \times 10^{5}$ & (8) \\
                0.76 & $5.58 \times 10^{5}$ &  \\
                0.76 & $1.97 \times 10^{5}$ &  \\
                0.65 & $5.57 \times 10^{4}$ &  \\
                0.62 & $2.19 \times 10^{4}$ &  \\
                0.55 & $8.06 \times 10^{3}$ &  \\
                0.55 & $8.19 \times 10^{3}$ & (9) \\
                2.19 & $4.56 \times 10^{6}$ & (10) \\
                1.25 & $2.96 \times 10^{6}$ &  \\
                0.88 & $1.39 \times 10^{6}$ &  \\
                0.69 & $1.54 \times 10^{5}$ &  \\
                0.55 & $8.19 \times 10^{3}$ &  \\
                0.44 & $1.50 \times 10^{3}$ &  \\
                0.55 & $1.21 \times 10^{4}$ & (11) \\
                0.55 & $8.19 \times 10^{3}$ & (12) \\
                0.55 & $1.21 \times 10^{4}$ &  \\
                0.55 & $8.19 \times 10^{3}$ &  \\
                0.55 & $8.81 \times 10^{3}$ &  \\
                0.55 & $2.19 \times 10^{4}$ &  \\
                0.55 & $8.73 \times 10^{3}$ & (13) \\
                160.00 & $(5.55 \pm 1.15) \times 10^{3}$ & (14) \\
                140.00 & $(9.31 \pm 0.83) \times 10^{3}$ &  \\
                90.00 & $(3.50 \pm 0.14) \times 10^{4}$ &  \\
                65.00 & $(6.39 \pm 0.63) \times 10^{4}$ &  \\
                0.23 & $2.68 \pm 0.03$ & (15) \\
                0.15 & $(2.77 \pm 0.12) \times 10^{-1}$ &  \\
                0.62 & $1.63 \times 10^{4}$ & (16) \\
                0.55 & $7.38 \times 10^{3}$ &  \\
                0.48 & $2.74 \times 10^{3}$ &  \\
                0.44 & $1.10 \times 10^{3}$ &  \\
                0.55 & $1.10 \times 10^{4}$ & (17) \\
                0.55 & $1.10 \times 10^{4}$ &  \\
                101.95 & $2.33 \times 10^{4}$ & (18) \\
                61.85 & $7.73 \times 10^{4}$ &  \\
                23.88 & $4.37 \times 10^{5}$ &  \\
                11.59 & $8.90 \times 10^{5}$ &  \\
                11.59 & $9.08 \times 10^{5}$ &  \\
                0.55 & $8.19 \times 10^{3}$ & (19) \\
                0.44 & $(1.95 \pm 0.02) \times 10^{3}$ & (20) \\
                \hline
        \end{tabular}
\end{table}

\setcounter{table}{2}

\begin{table}
        \centering
        \caption{Continued.}
        \begin{tabular}{lll}
                \hline\hline
                Wavelength ($\mu$m) & Photometry (mJy) & References \\
                \hline          
                4.89 & $(1.33 \pm 0.03) \times 10^{6}$ & (21) \\
                3.52 & $(2.89 \pm 0.14) \times 10^{6}$ &  \\
                2.22 & $(4.44 \pm 0.17) \times 10^{6}$ &  \\
                1.26 & $(3.25 \pm 0.22) \times 10^{6}$ &  \\
                160.00 & $5.19 \times 10^{3}$ & (22) \\
                140.00 & $7.66 \times 10^{3}$ &  \\
                90.00 & $3.47 \times 10^{4}$ &  \\
                65.00 & $3.12 \times 10^{4}$ &  \\
                61.85 & $7.73 \times 10^{4}$ & (23) \\
                23.88 & $4.37 \times 10^{5}$ &  \\
                22.09 & $5.76 \times 10^{5}$ &  \\
                11.59 & $9.08 \times 10^{5}$ &  \\
                2.19 & $(5.69 \pm 1.47) \times 10^{6}$ &  \\
                2.19 & $4.21 \times 10^{6}$ & (24) \\
                101.95 & $2.39 \times 10^{4}$ &  \\
                61.85 & $8.87 \times 10^{4}$ &  \\
                23.88 & $4.72 \times 10^{5}$ &  \\
                11.59 & $9.69 \times 10^{5}$ &  \\
                101.95 & $2.58 \times 10^{4}$ &  \\
                61.85 & $9.22 \times 10^{4}$ &  \\
                23.88 & $4.19 \times 10^{5}$ &  \\
                11.59 & $1.42 \times 10^{6}$ &  \\
                0.53 & $(1.03 \pm 0.01) \times 10^{4}$ & (25) \\
                0.42 & $(8.55 \pm 0.16) \times 10^{2}$ &  \\
                2.19 & $4.73 \times 10^{6}$ & (26) \\
                0.55 & $1.25 \times 10^{4}$ & (27) \\
                0.44 & $1.40 \times 10^{3}$ &  \\
                0.53 & $(1.03 \pm 0.01) \times 10^{4}$ & (28) \\
                0.42 & $(8.56 \pm 0.14) \times 10^{2}$ &  \\    
                0.33 & $5.86 \times 10^{1}$ & (29, 31) \\       
                0.35 & $8.72 \times 10^{1}$ & \\
                0.37 & $1.13 \times 10^{2}$ & \\
                0.45 & $2.08 \times 10^{3}$ & \\
                0.52 & $5.16 \times 10^{3}$ & \\
                0.36 & $5.06 \times 10^{1}$ & (30, 31) \\
                0.45 & $1.71 \times 10^{3}$ & \\
                0.55 & $8.09 \times 10^{3}$ & \\
                \hline
        \end{tabular}
        \tablefoot{For the line with blank references, the previous reference applies.}
        \tablebib{
                (1)~\citetads{1997ESASP1200.....E}; 
                (2)~\citetads{1994AJ....107..280H}; 
                (3)~\citetads{1998AJ....115.1212U}; 
                (4)~\citetads{2002A&A...384..180F}; 
                (5)~\citetads{2008AJ....136..735L}; 
                (6)~\citetads{2011AJ....142...15G}; 
                (7)~\citetads{2016A&A...595A...1G, 2021A&A...649A...1G}; 
                (8)~\citetads{2016A&A...595A...1G,2023A&A...674A...1G}; 
                (9)~\citetads{1966CoLPL...4...99J}; 
                (10)~\citetads{1978A&AS...34..477M}; 
                (11)~\citetads{1975A&AS...22..239N}; 
                (12)~\citetads{1975A&AS...22..239N}; 
                (13)~\citetads{2001ApJ...558..309D}; 
                (14)~\citetads{2009ASPC..418....3Y}; 
                (15)~\citetads{2017ApJS..230...24B}; 
                (16)~\citetads{2015AAS...22533616H}; 
                (17)~\citetads{1985AbaOB..59...83B}; 
                (18)~\citetads{1996AJ....112.2862E}; 
                (19)~\citetads{1879RNAO....1....1G};
                (20)~\citetads{ 2015yCat.5145....0M}; 
                (21)~\citetads{2010ApJS..190..203P}; 
                (22)~\citetads{2010A&A...514A...3P}; 
                (23)~\citetads{2019A&A...622A.120U}; 
                (24)~\citetads{1998A&AS..129..363J}; 
                (25)~\citetads{2008PASP..120.1128O}; 
                (26)~\citetads{2009MNRAS.400.1945T}; 
                (27)~\citetads{2016MNRAS.463.4210N};  
                (28)~\citetads{2011AstL...37..707G};
                (29)~\citetads{1975RMxAA...1..299J};
                (30)~\citetads{1988csmg.book.....R};
                (31)~\citetads{2024RASTI...3...89M}.
        }
\end{table}

\FloatBarrier

\section{Details of the SPHERE observations}

\begin{figure*}
        \centering
        \includegraphics[width=\textwidth]{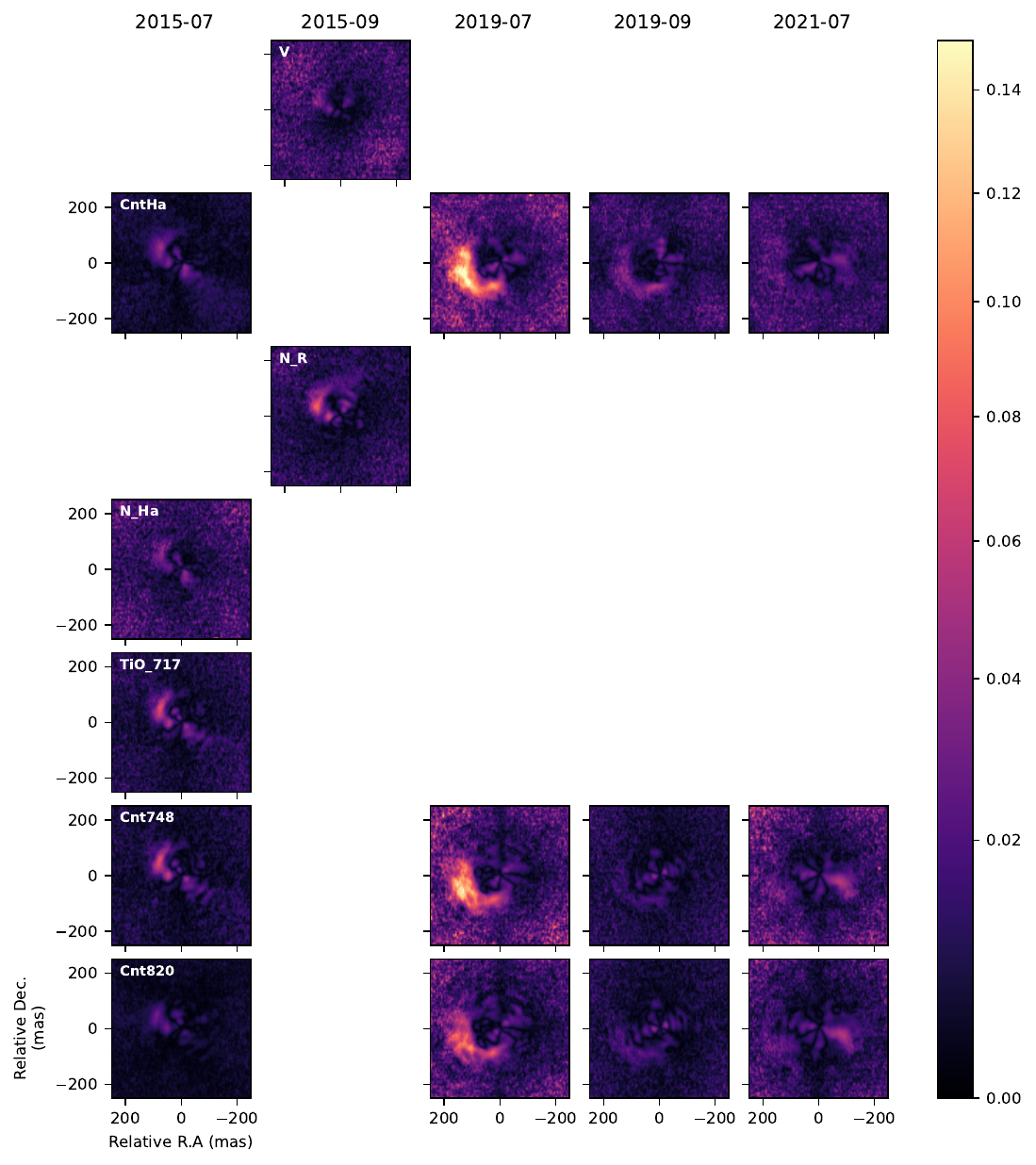}
        \caption{Degree of linear polarization observed by VLT/SPHERE-ZIMPOL on $\pi^1$~Gru. The (0, 0) origin of each image corresponds to the position of the AGB star from the intensity frames (Stokes $I$). Each column corresponds to a different epoch. Each row corresponds to a ZIMPOL filter.}\label{Fig:all_DoLP}
\end{figure*}

\begin{figure*}
        \centering
        \includegraphics[width=\textwidth]{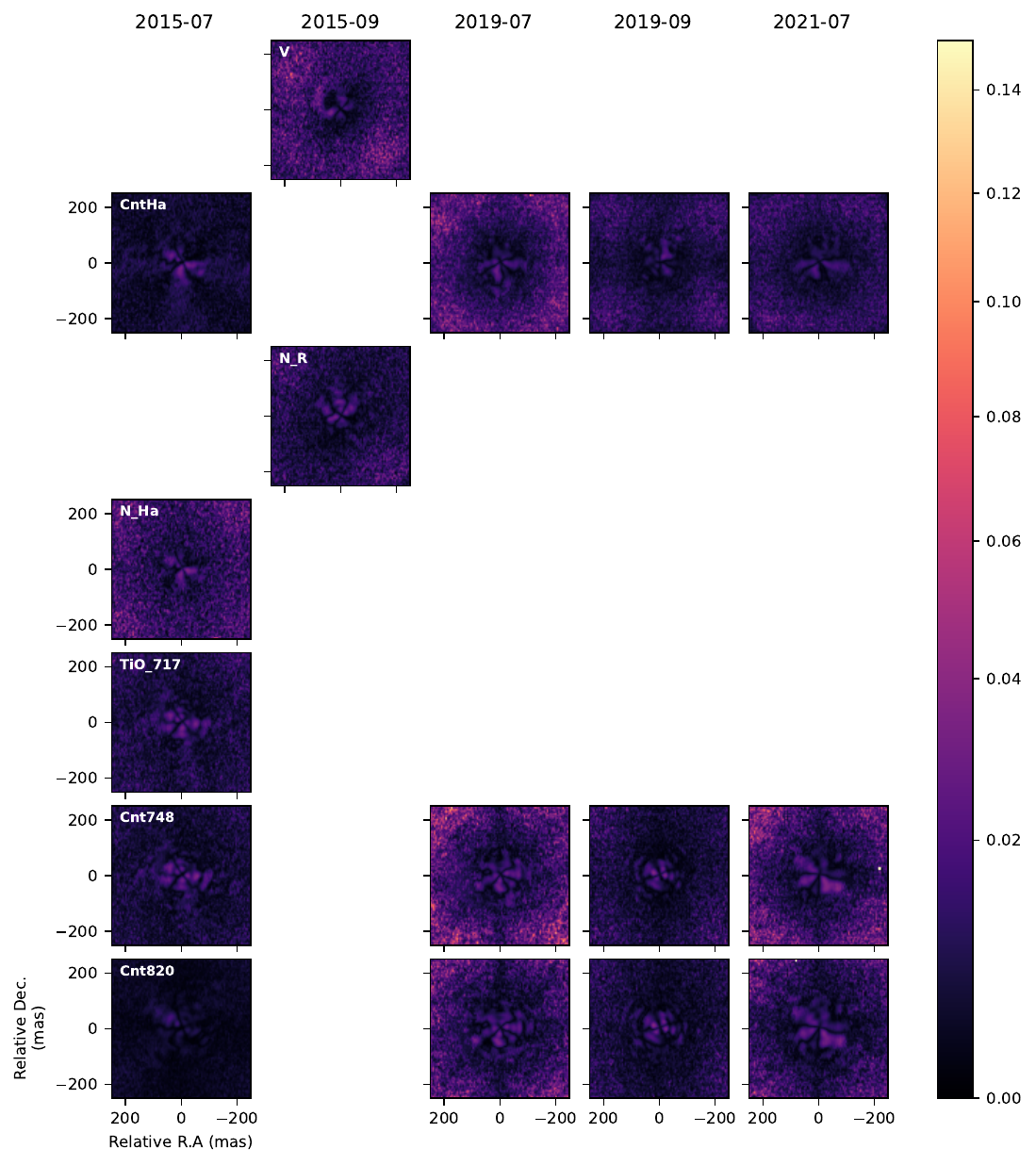}
        \caption{Degree of linear polarization rms observed by VLT/SPHERE-ZIMPOL on $\pi^1$~Gru. The (0, 0) origin of each image corresponds to the position of the AGB star from the intensity frames (Stokes $I$). Each column corresponds to a different epoch. Each line corresponds to a ZIMPOL filter.}\label{Fig:all_DoLP_rms}
\end{figure*}

Figure~\ref{Fig:all_DoLP} (resp. Fig.~\ref{Fig:all_DoLP_rms}) shows the complete set of observations of the $\pi^1$~Gru system with VLT/SPHERE-ZIMPOL for the DoLP (resp. DoLP rms).

\FloatBarrier

\section{True anomaly of the best fitted orbital solution at observed epochs}

\begin{table}[!ht]
        \centering
        \caption{True anomaly for the best orbital solution of Sect.~\ref{SubSect:Orbit}, derived at the observation epochs of ALMA, and ZIMPOL.\label{Tab:TrueAnom}}
        \begin{tabular}{l l l}
                \hline\hline
                \noalign{\smallskip}
                Epoch (yr) & $\theta$ ($^{\circ}$) & Instrument\\
                \hline
                \noalign{\smallskip}
                2015.52 & 31.6 & ZIMPOL \\
                2015.73 & 44.7 & ZIMPOL \\
                2019.48 & 159.7 & ALMA \\
                2019.52 & 160.4 & ZIMPOL \\
                2021.51 & 193.5 & ZIMPOL \\
                \hline
        \end{tabular}
\end{table}

\end{appendix}

\end{document}